\documentclass[fleqn,usenatbib]{mnras}

\usepackage{newtxtext,newtxmath}
\usepackage[T1]{fontenc}
\usepackage{graphicx} 
\usepackage{amsmath} 
\usepackage{comment}
\usepackage{tikz}\usetikzlibrary{shapes.geometric, arrows}
\usepackage{pgfplots}\pgfplotsset{compat=1.18}
\usepackage{pifont}
\usepackage{tikz}
\usetikzlibrary{arrows.meta}
\tikzset{%
  >={Latex[width=2mm,length=2mm]},
    base/.style = {
        rectangle, rounded corners, draw=black,
        minimum width=2cm, minimum height=1cm,
        text centered, font=\sffamily
    }, 
    blind/.style = {base, fill=blue!30},
    intermediate/.style = {base, fill=orange!15},
    discarded/.style = {base, fill=red!60},
    verified/.style = {base, fill=green!60}
}

\DeclareUnicodeCharacter{2212}{-}
\DeclareRobustCommand{\VAN}[3]{#2}
\let\VANthebibliography\thebibliography
\def\thebibliography{\DeclareRobustCommand{\VAN}[3]{##3}\VANthebibliography}

\usepackage[utf8]{inputenc}
\usepackage{blindtext}
\usepackage{color}
\usepackage{everysel}
\usepackage{fancyhdr} 
\usepackage{float}
\usepackage{physics}
\usepackage{textcomp}
\usepackage{clrscode}
\usepackage{verbatim}
\usepackage[none]{hyphenat} 
\usepackage{amsfonts} 
\usepackage{caption}
\usepackage{subcaption}
\usepackage{siunitx}
\usepackage{mathtools}
\usepackage{lipsum}
\usepackage{tgcursor}
\usepackage{cases}
\usepackage{multirow}
\usepackage{ragged2e}

\usepackage{xcolor}
\usepackage{longtable}
\usepackage{booktabs}
\usepackage{graphicx} 
\usepackage{booktabs} 
\usepackage{xparse} 
\usepackage{supertabular}
\usepackage{makecell}

\usepackage{array}
\newcolumntype{x}[1]{>{\centering\arraybackslash\hspace{0pt}}p{#1}}

\newif\ifsupp
\supptrue

\newif\ifbf
\bffalse

\newcommand{\mybf}[1]{\ifbf {\bf \textcolor{magenta}{#1}}\else #1\fi}

\usepackage{hyperref}
\hypersetup{
    colorlinks=true,
    linkcolor=blue,
    filecolor=magenta,      
    urlcolor=cyan,
}

\newcommand{\namedref}[2]{\hyperref[#2]{#1~\ref*{#2}}\xspace}
\newcommand{\figureref}[1]{\namedref{Figure}{fig:#1}}
\newcommand{\tableref}[1]{\namedref{Table}{table:#1}}

\newcommand{\sectionref}[1]{\namedref{Section}{sec:#1}}

\newcommand{\sersic}{S\'{e}rsic\xspace}
\newcommand{\ie}{{\textit{i.e.,}}\xspace}
\newcommand{\objid}{{\texttt{objID}}\xspace}
\newcommand{\objids}{{\texttt{objID}s}\xspace}
\newcommand{\specobjid}{{\texttt{specObjID}}\xspace}
\newcommand{\specobjids}{{\texttt{specObjID}s}\xspace}
\newcommand{\gothic}{{\textsc{Gothic}}\xspace}

\title[Automated Detection of Double Nuclei Galaxies and Discovery of Dual AGN]{Automated Detection of Double Nuclei Galaxies using \gothic and the Discovery of a Large Sample of Dual AGN}

\author[A. Bhattacharya et al.]{\parbox{\linewidth}{
Anwesh Bhattacharya$^{1}$, Nehal C. P.$^{2}$, Mousumi Das$^{3}$, Abhishek Paswan$^{3,4}$, Snehanshu Saha$^{5}$\thanks{Corresponding author - snehanshu.saha@ieee.org},\\ Fran\c{c}oise Combes$^{6}$}\\
\\
$^{1}$Dept. of Physics and CSIS, Birla Institute of Technology \& Science, Pilani, India\\
$^{2}$Dept. of Physics, Indian Institute of Science Education and Research, Bhopal\\
$^{3}$Indian Institute of Astrophysics, Bangalore\\
$^{4}$ Department of Physics, University of Allahabad, India\\
$^{5}$ APPCAIR, Dept. of CSIS, Birla Institute of Technology \& Science, Goa, and HappyMonk AI, India\\
$^{6}$ Observatoire de Paris, LERMA, Coll\`ege de France, PSL University, Sorbonne University, CNRS, Paris, France}

\date{Accepted 2023 July 10. Received 2023 July 10; in original form 2022 November 16}

\pubyear{2022}

\begin{document}
\label{firstpage}
\pagerange{\pageref{firstpage}--\pageref{lastpage}}

\maketitle

\begin{abstract}
We present a novel algorithm to detect double nuclei galaxies (DNG) called \gothic (\textbf{G}raph-B\textbf{o}os\textbf{t}ed iterated \textbf{H}\textbf{i}ll \textbf{C}limbing) — that detects whether a given image of a galaxy has two or more closely separated nuclei. Our aim is to test for the presence of dual/multiple active galactic nuclei (AGN) in galaxies that visually represent a DNG. Although galaxy mergers are common, the detection of dual AGN is rare. Their detection is very important as they help us understand the formation of supermassive black hole (SMBH) binaries, SMBH growth and AGN feedback effects in multiple nuclei systems. There is thus a need for an algorithm to do a systematic survey of existing imaging data for the discovery of DNGs and dual AGNs. We have tested \gothic on an established sample of DNGs with $100\%$ detection rate and subsequently a blind search of 1 million SDSS DR16 galaxies (with spectroscopic data available) lying in the redshift range of $z=0$ to $0.75$. From the list of candidate DNGs found, we have detected 159 dual AGNs, of which 2 are triple AGN systems. Our results show that dual AGNs are not common, and triple AGN even rarer. The color (u-r) magnitude plots of the DNGs indicate that star formation is quenched as the nuclei come closer and as the AGN fraction increases. The quenching is especially prominent for dual/triple AGN galaxies that lie in the extreme end of the red sequence. 
\end{abstract}


\section{Introduction}

Galaxy mergers are the main mechanism for the hierarchical growth of galaxies in our universe \citep{white.rees.1978,patton.etal.2020}. Depending upon the gas content of galaxies, gas rich mergers can lead to substantial amounts of star formation, resulting in starbursts and ultraluminous infrared galaxies (ULIRG hereafter) \citep{mirabel.sanders.1996}. Dry or gas free mergers of elliptical and S0 galaxies do not result in star formation but lead to the growth of more massive early type galaxies \citep{thomas.etal.2005}. As galaxies merge, the bulges and nuclei of the individual galaxies come closer leading to the formation of dual or even multiple nuclei systems that are often enclosed within a single galaxy envelope \citep{capelo.etal.2017}. In the process, one or more of the supermassive black holes (SMBH hereafter) in the galaxy nuclei may start accreting mass leading to the formation of active galactic nuclei (AGN hereafter) \citep{shlosma.etal.1990,peterson.1997}. Alternatively, large amounts of gas may fall into the nuclear regions of the merging galaxies triggering nuclear starburst activity \citep{kim.etal.2021}. 


Thus, double nuclei galaxies (DNG hereafter) can host AGN pairs, AGN-starburst pairs or individual star forming nuclei in their centers \mybf{\citep{rubinur.etal.2019,DeRosa.etal.2019}}. When two or more AGN are found in a merger remnant and are separated by $<$10 kpc, it is often termed as a dual AGN (DAGN hereafter) system, although the definition varies in the literature and can include AGN at separations of upto $\sim50 kpc$ \citep{Koss_2012}. Triple AGN systems have also been detected in small groups of merging galaxies \citep{yadav.etal.2021} which suggests that multiple AGN systems are more common than we think. In fact simulations have shown that minor mergers can lead to many massive black holes (MBHs) that are spatially offset from the central galaxy nucleus; such MBHs are called "wandering black holes", since they do not appear to be in the process of merging with the nuclear SMBH \citep{bellovary.etal.2010,ricarte.etal.2021}. A few wandering MBHs have been detected \citep{greene.etal.2021,reines.etal.2020}. If an AGN pair is separated by only a few parsecs or $<$1 kpc, it is termed as a binary AGN (BAGN). The number of DAGN identified till now is  $<$100 \citep{rubinur.etal.2018,stemo.etal.2021} and the number of \mybf{confirmed binary AGN is $<$ 10 \citep{kharb.etal.2017a,ciurlo.etal.2023}}. These numbers are small compared to the total fraction of AGN detected in galaxies at low redshifts. Since galaxy mergers are common, this suggests that there is a large population of dual/multiple AGN systems that have not yet been detected.

Most of the early detections of DAGN in the literature were serendipitous, and were discovered during galaxy surveys of emission line galaxies and merger remnants \citep{komossa.etal.2003} or radio observations of compact sources \mybf{\citep{DeRosa.etal.2019}}. Later studies used double peaked AGN emission line galaxies (DPAGN hereafter) to search for DAGN \citep{zhou.etal.2004,rubinur.etal.2019,maschmann.etal.2020}. However, DPAGN can also be due to AGN outflows or rotating disks \citep{kharb.etal.2017b}. Recent studies have shown that one of the best ways to detect DAGN is by using a parent sample of dual nuclei galaxies and then identifying the nature of their nuclei using multi-wavelength observations \citep{DeRosa.etal.2019,scialpi.etal.2023,koss.etal.2023}. However, the latest surveys of galaxy pairs are either too small or the nuclei separations are too large to be defined as dual nuclei systems \citep{patton.etal.2016}. 

There are several reasons why we need to find more dual or multiple AGN systems. One of the most important reasons is that they represent the first stage in the formation of SMBHs pairs. At distances of several Mpc beyond the local universe, the only way to detect SMBHs is via the radiation emitted from AGN activity. Although the gravitational radiation from SMBH pairs is emitted only when they are separated by less than a parsec, a large sample of DAGN will help us understand how they form and evolve. So revealing the frequency and nature of SMBH pairs is the first step in understanding the low frequency (nano Hz) gravitational waves radiated by inspiralling SMBHs \citep{aggarwal.etal.2019,goulding.etal.2019,shannon.etal.2015}. Secondly, although dual AGN/multiple AGN are rare, it has been shown that galaxy mergers are important triggers of AGN activity and the formation of DAGN, both of which can lead to nuclear star formation and SMBH growth. Simulations have suggested that merger-triggered AGN may dominate SMBH growth \citep{hopkins}. Finally, SMBH pairs or clusters will affect the AGN feedback and stellar winds/outflows associated with galaxy mergers, which can result in the growth of the circumgalactic medium (CGM hereafter) around galaxies \citep{fielding.etal.2020,davies.etal.2020}. Multiple AGN can give rise to stronger winds and outflows, producing additional feedback effects as observed in nearby DAGN \citep{mazzarella.etal.2012,rubinur.mrk212.2020}. Hence, large samples of DAGN are essential for fully understanding the role of mergers in SMBH growth and galaxy evolution.

In this paper we present an automated search for dual/multiple nuclei in merging and interacting galaxies, using an algorithm called \gothic that detects such systems using optical images. We applied \gothic to SDSS images and derived a large sample of double and multiple nuclei in closely interacting galaxies and merger remnants. Our broad goal \mybf{is to} create a large sample of dual or multiple nuclei from which we can detect DAGN (or even multiple AGN systems). In \sectionref{motivation} we first review previous studies that search for galaxy pairs and then build the case for the need of a large catalog of DNGs. We also discuss the trial sample that was used as a benchmark for developing the algorithm. In \sectionref{gothic}, we describe the design of the algorithm. The algorithm has been carefully designed such that it achieves a $100\%$ detection rate on the trial sample, whilst keeping the false positive rate reasonably low on previously unseen samples.

We subsequently performed a blind search by applying the algorithm on a large sample of 1 million galaxies and detected $\approx100,000$ candidate DNGs. We filtered these candidates and manually analyzed a smaller subset of them in \sectionref{selection} and \sectionref{analysis}. We then present the DNGs in a catalogue and discuss their properties. We have separated out the dual and multiple AGN in the sample and discussed the nature of the host galaxies. Finally, we conclude with a discussion of the impact of our study on understanding DAGN and galaxy mergers in \sectionref{discussion} and \sectionref{conclusion}. 

\section{Why Automate the search of Double Nuclei Galaxies?}
\label{sec:motivation}

\subsection{Previous Studies and our Motivation}
Large samples of galaxy pairs have been previously derived in the literature from imaging or spectroscopic surveys, by selecting the nearest neighbor galaxies. This was usually done by constructing radial bins around galaxies and selecting neighboring galaxies that lie within specific radii from the target galaxy (e.g. $R<$2 Mpc) \citep{busmante.etal.2020}. Alternatively, differences in systemic velocities between a target galaxy and its neighbors \mybf{have been} used to derive the closest neighbors \citep{maschmann.etal.2020}. Although this method did select galaxy pairs, it was mainly aimed at detecting the tidal features in interacting galaxies \citep{ellison.etal.2010} and the effect of galaxy interactions on star formation \citep{barton.etal.2000,patton.atfiedld.2008}. Most of the galaxy pairs that were detected in these surveys were separated by several megaparsec (Mpc) distances, and hence these samples basically represent galaxies interacting at a distance but not actually merging. Many of these surveys also have constraints on the mass ratios of the two galaxies and contain distant pairs and not close galaxy pairs \citep{busmante.etal.2020}. These methods are thus biased towards detecting galaxies interacting at a distance rather than merger remnants that contain two or more nuclei encased in a single envelope \mybf{that represent} closely interacting galaxies. 

There have recently been several studies of galaxy pairs in cosmological simulations. These studies use hydrodynamical simulations to emulate galaxy mergers which are then used to build mock samples upon which an image-classification scheme is developed. However, some studies suggest that mock images may result in poor performance in merger classifications \citep{deep.learning.merger}. Others have performed convolutional neural network (CNN) training on processed isolated galaxy images (e.g. IllustrisTNG) for the task of automated merger classification \citep{cnn.merger,illustris}. They obtain a sample with a purity of only $6\%$ due to the intrinsic rarity of galaxy mergers. Thus, there is clearly some difficulty in building a clean sample of galaxy mergers using simulations \citep{nevin}. Studies have also compared the efficiency of using stellar kinematics versus visual images and found that visual identification is better \citep{bottrell.etal.2021}. This is clearly demonstrated in \citep{visual.identification} where samples of \emph{Visually Identified Pairs} (VIP) are presented using simulated images from IllustrisTNG.

Thus there is a need for an algorithm and pipeline that can sift through large optical or near-infrared (NIR) imaging surveys of real galaxies and extract double/multiple galaxy nuclei efficiently with no bias with respect to spatial separations, galaxy mass ratios or redshift ranges of the galaxies. Such a method will be computationally more efficient compared to counting nearest neighbors in radial bins. It is with this aim in mind that we have developed an algorithm that can detect dual nuclei systems at kiloparsec (kpc) scale separations from an optical survey and identify a pool of dual nuclei systems that can be used to study galaxy mergers and detect candidate dual/multiple AGN systems. In the following section we discuss the trial sample of dual nuclei that we have used in our study.

\subsection{The Trial Sample}
Since DNGs are a rare type of galactic object, it is essential to have a sample of established DNGs which can be used as a benchmark to check the performance of our algorithm. 
The Gimeno Catalog \citep{Gimeno_2004} presents a compiled sample of 107 DNGs from previously existing catalogs. They have also used ESO plates to catalogue the sources. However, we found that only 47 galaxies out of the 107 had valid entries in SDSS DR16. Thus, we used these 47 galaxies as the trial sample.


\begin{figure}
    \centering
    \includegraphics{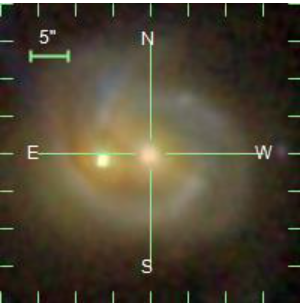}
    \caption{Color composite SDSS image of MRK 739 showing clearly the two nuclei.}
    \label{fig:ngc3758.sdss}
\end{figure}

MRK 739 (\figureref{ngc3758.sdss}) is a prototypical example of the type of galaxy we are interested in detecting. It was first studied in detail by \citet{ngc3758} where it was confirmed to host a DAGN. It was more recently studied by \citet{ngc3758.recent} who identified the western and eastern nuclei to be on their ongoing first passage. The construction of our algorithm is targeted towards finding galaxies that resemble the visual features of MRK 739 \ie there should be two distinct bulges. 

\mybf{
The GC catalog was subsequently used by \citet{Mezcua_2014} in which they selectively studied  DNGs that were available in SDSS DR8. The study focuses on classifying the DNG sources as major/minor mergers and to find evidence for AGN activity in the DNGs. This is because DNGs are a strong candidate for DAGN systems and detecting them accurately can help in building a catalog of DAGNs. In fact, there are numerous studies showing that visual DNG systems turn out to be hosts for DAGNs \citep{rubinur.mrk212.2020, Koss_2012, komossa.etal.2003, koss.recent.1, koss.recent.2}. 
}

\section{Description of the Procedure}
\label{sec:gothic}

To determine whether a galaxy is a DNG we need to first have image cutouts of galaxies from SDSS. 
Figure \ref{fig:ngc3758.sdss} is an example of a color composite image of a DNG in the SDSS display in object explorer\footnote{
\mybf{\url{http://skyserver.sdss.org/dr16/en/tools/explore/summary.aspx}}
}. It is a $40'' \times 40''$ square-cutout centred on the coordinates with which the \objid of the galaxy is associated. A successful algorithm for DNG detection must be able to identify the light envelope of one galaxy or two close galaxies, and detect the presence of two distinct bulges if any. In practice, however, \objid of galaxies in SDSS do not always faithfully correspond to a galaxy \ie some IDs correspond to stars, QSOs, or blank patches of sky with no visible object in its vicinity. So the detection procedure must carefully handle mislabelled galaxies. This unreliable nature of the \emph{ground truth} of SDSS labels has been previously \mybf{documented} \citep{star.quasar}.

In terms of its functionality, the \gothic algorithm takes as input a $40'' \times 40''$ square cutout of a galaxy and classifies it as a DNG or a single-nuclei galaxy. In a nutshell, this classification is done by searching for two well-defined peaks of light that are characteristic of DNGs. The algorithm consists of six major components which are briefly summarized below
\begin{enumerate}
    \item \textbf{Image Normalization and Smoothing} - The first step of the algorithm is to smooth the input image so that it is amenable to further processing
    \item \textbf{Edge Detection} - The galaxy of interest occupies a portion of the $40`` \times 40``$ cutout space. To visually identify the boundaries of the galaxy, we use the Canny edge detector \citep{canny}. 
    \item \textbf{Fitting the Light Profile} - After the galaxy has been located in the cutout, we compute a histogram of the pixel values contained in the galaxy. This histogram is then fit to a \mybf{\sersic} light profile \citep{sersic} to determine the \mybf{\sersic} index.
    \item \textbf{Determination of the Search Region} - After determining the exact \mybf{\sersic} profile, a pixel \emph{cutoff value} is calculated. Only the pixels above this cutoff value are the ones with appreciable light intensity. The search for light peaks occurs in the region above this cutoff value.
    \item \textbf{Iterated Hill Climbing} - This is a formal algorithm for locating peaks of light in the search region determined by the previous step. 
    \item \textbf{Final Classification} - Typically, more than 2 peaks are detected for a given galaxy. This occurs since the cutout images are inherently noisy, even after smoothing. Using all the information determined by all previous steps, we develop a sound rule-based system to judge whether the peaks represent those of a DNG or not.
\end{enumerate}

Detailed descriptions of each of the above 6 steps can be found in Supplementary Section 1. 

\input{Tables/confusion}
\begin{figure*}
\begin{tikzpicture}[node distance=1.5cm,
    every node/.style={fill=white, font=\sffamily}, align=left]
  \node (blind)          [blind, text width=2cm]                                             {Blind sample of 1 million objects};
  \node (detections)     [intermediate, right of=blind, xshift=3.5cm]          {104,412 detections};
  \node (inconsistent)    [discarded, above of=detections, yshift=1cm]   {9,253 objects};
  \node (consistent)     [intermediate, below of=detections, yshift=-1cm]     {95,159 objects};
  \node (redshift)     [intermediate, right of=consistent, xshift=3.5cm]     {47,521 objects};
  \node (spec)     [intermediate, right of=redshift, xshift=3.5cm]     {949 objects};
  \node (close)     [discarded, above of=redshift, yshift=3.5cm]     {46,061 objects};
  \node (ver)     [verified, above of=spec, yshift=3.5cm]     {681 objects};
       
  \draw[->]       (blind) -- node[text width=2cm, anchor=south]{\scriptsize{Detected by \gothic ($~10\%$)}} (detections);
  \draw[->]       (detections.north) -- node[text width=2cm, anchor=west]
                                   {\scriptsize{\textbf{1 -} Discarded due to inconsistent dual peaks ($~9\%$)}}
                                   (inconsistent);
  \draw[->]       (detections.south) -- node[text width=2cm, anchor=west]
                                   {\scriptsize{\textbf{1 -} Dual peaks detected consistently across bands ($~91\%$)}}
                                   (consistent);
  \draw[->]       (consistent) -- node[text width=2cm, xshift=0.15cm, anchor=south]
                                   {\scriptsize{\textbf{2 -} \specobjid available ($~50\%$)}}
                                   (redshift);
  \draw[->]       (redshift) -- node[yshift=0.1cm, text width=2.5cm, anchor=south]
                                   {\scriptsize{\textbf{3 -} Nuclei with distinct \specobjid and $\Delta d<40$kpc ($~2\%$)}}
                                   (spec);
  \draw[->]       (redshift.north) -- node[yshift=0.1cm, xshift=0.1cm, text width=2cm, anchor=west]
                                   {\scriptsize{\textbf{3 -} Both nuclei in fibre ($~97\%$)}}
                                   (close);
  \draw[->]       (spec.north) -- node[yshift=0.1cm, xshift=0.1cm, text width=2cm, anchor=west]
                                   {\scriptsize{Manual verification ($~71\%$)}}
                                   (ver);
  
\end{tikzpicture}
\caption{Flowchart demonstrating the filtration of the blind sample down to the visually confirmed sample. The bold numbers refer to the filtration steps in \sectionref{selection} and the numbers in parenthesis indicate the percentage reduction in the sample size caused by each filtration step}
\label{fig:filtration.flowchart}
\end{figure*}

\subsection{Testing the performance of \gothic}

Having developed \gothic, we tested its performance on a test sample that comprises the following
\begin{enumerate}
    \item \textbf{Positive sample} - All the 47 galaxies in the trial sample serve as the positive sample.
    \item \textbf{Negative control sample} - We randomly selected 47 single nucleus galaxies from SDSS DR16. We ensured the size of control sample was equal to the positive sample for fairness.
\end{enumerate}

We found that \gothic had a $100\%$ detection rate on the positive sample \ie all DNGs were correctly recognized and there were 0 false negatives. On the other hand, only 7 galaxies from the control sample were incorrectly recognized as DNGs. This represents a false positive rate of $~15\%$. To visualize the various detections in a concise format, we provide a concise confusion matrix in \tableref{confusion}. The details of \gothic's performance on the test sample can be found in Supplementary Section 2.

\section{Applying \gothic to a Large Blind Sample}
\label{sec:selection}

As stated previously in \sectionref{motivation}, our goal is to prepare a sample of candidate DAGNs. To do so, we first need to identify galaxies that visually resemble DNGs and subsequently analyse their spectra to confirm if it is truly an AGN. Given \gothic's $100\%$ detection rate on the Gimeno Catalog, we can expect that if the image given to it were that of a DAGN, it would recognize it as a DNG with high certainty. Thus, if it is run upon a large sample of randomly chosen galaxies (most of which are bound to be single nucleus galaxies), it will filter out those rare galaxies that resemble DNGs. To that end, we ran \gothic on 1 million galaxies chosen uniformly at random from SDSS DR16 that have spectroscopic data available. The galaxies were chosen uniformly at random so that there is no bias towards any physical property (such as redshift or magnitude in any band). We used images in the $u,g,r,i$ bands and discarded the $z$ band as we found images in this band to be often noisy. Among the 1 million randomly chosen galaxies, \gothic detected 104,412 DNGs which represents an overall hit rate of $\approx 10\%$. However, using these \emph{raw} detections for further analysis is not suitable due to the following reasons
\begin{enumerate}
    \item Extrapolating from \gothic's performance on the trial sample, we can expect false positives in the detections which need to be filtered out. 
    \item The raw detection sample is too large to be analysed manually.
\end{enumerate}

\subsection{Filtration Steps to Obtain a Reduced Sample}
\label{sec:filtration}

Thus we need to filter the raw detections such that --- i) Detections that are likely to be false positives are removed. ii) The reduced sample will only contain those detections that are highly likely to be DAGN after manual analysis. Thus we applied the following sequence of conditions to arrive at a smaller sample that could be manually studied (depicted as a flowchart in \figureref{filtration.flowchart}). 

\noindent\textbf{Filtration Step 1} - The pixel coordinates of the detected nuclei need to be consistent across all the bands in which they are detected in. For example, if a peak is detected at $(x,y)$ in the $g$ band and \gothic reports a double detection in the $r$ band, a corresponding peak has to be present in the $r$ band within the $3\times3$ window centred at $(x,y)$. If a given galaxy has a consistent detection of peaks across all bands in which \gothic reports a detection, we denote it as a \emph{pure double}. Otherwise, we denote it as an \emph{impure double} and intuitively these galaxies are more likely to be false positives. We found $95,159$ pure doubles and used this subset for further filtration. The remaining $9,353$ impure doubles were exempt from further study.\\

\noindent\textbf{Filtration Step 2} - Although our initial one million sample had spectroscopic data available, it is not necessarily true that the DNG detected by \gothic always correspond to the galaxy referenced by its \objid. \gothic is purely an image processing algorithm and is unaware of the mapping between galaxies/peaks and \objids. Thus, we performed a coordinate to \objid lookup of the DNGs detected by \gothic. We found that $47,521$ of the $95,159$ pure doubles have spectroscopic data available in SDSS (indicated by an existing \specobjid entry). Spectroscopic information is needed in our final analysis (redshift distribution, for example) and thus this step is necessary.\\
    
\noindent\textbf{Filtration Step 3} - Of this filtered spectroscopic DNG sample, we choose only those DNGs that have distinct \specobjids corresponding to \emph{both} the peaks of the DNGs. So from the $47,521$ candidate DNGs, we obtained two groups of sources --- i) The first was composed of 46,061 DNG candidates where both nuclei were located within the SDSS fibre which has a diameter 3$^{\prime\prime}$. These sources could not be distinguished as two spectroscopic sources and hence discarded. ii) The second group is a sample of 949 DNGs where both nuclei have distinct \specobjids and are very close in redshift (z). The cutoff for $\Delta$z  separation is defined by a distance (d) which is taken to be approximately $\Delta d<40$kpc, for all redshifts. This is the harshest filtration step as it retains only \(2\%\) of the previous sample. If SDSS catalogues distinct \specobjids at the location of the peaks, we gain additional confidence that the detection is a true positive.\\




\begin{figure*}
    \centering
    \begin{subfigure}[b]{0.5\textwidth}
        \hspace{12pt}
        \includegraphics[width=0.85\textwidth,height=0.15\textheight]{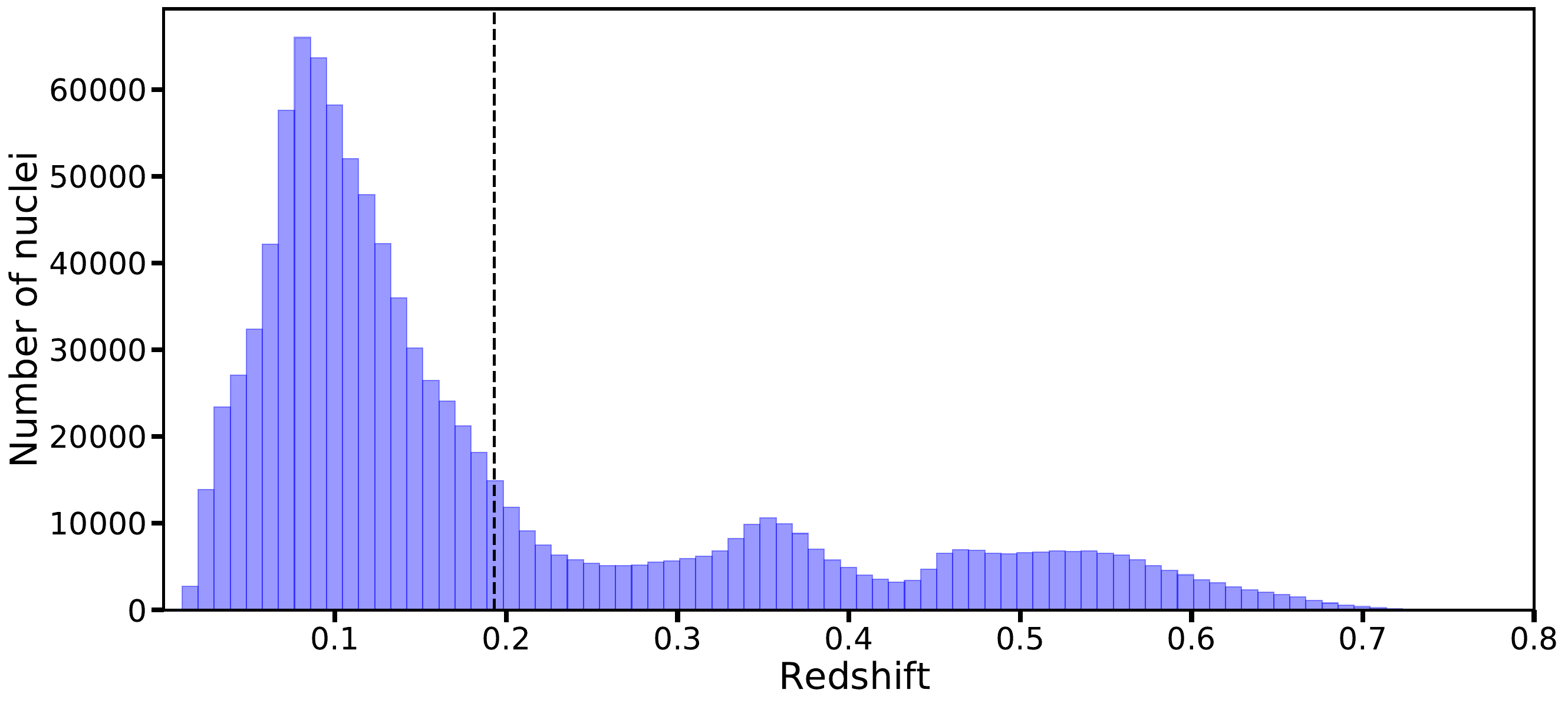}
        \vspace{-5pt}
        \caption{Starting 1 million sample}
        \label{fig:redshift.mill}
    \end{subfigure}%
    \begin{subfigure}[b]{0.5\textwidth}
        \vspace{10pt}\hspace{12pt}
        \includegraphics[width=0.85\textwidth,height=0.15\textheight]{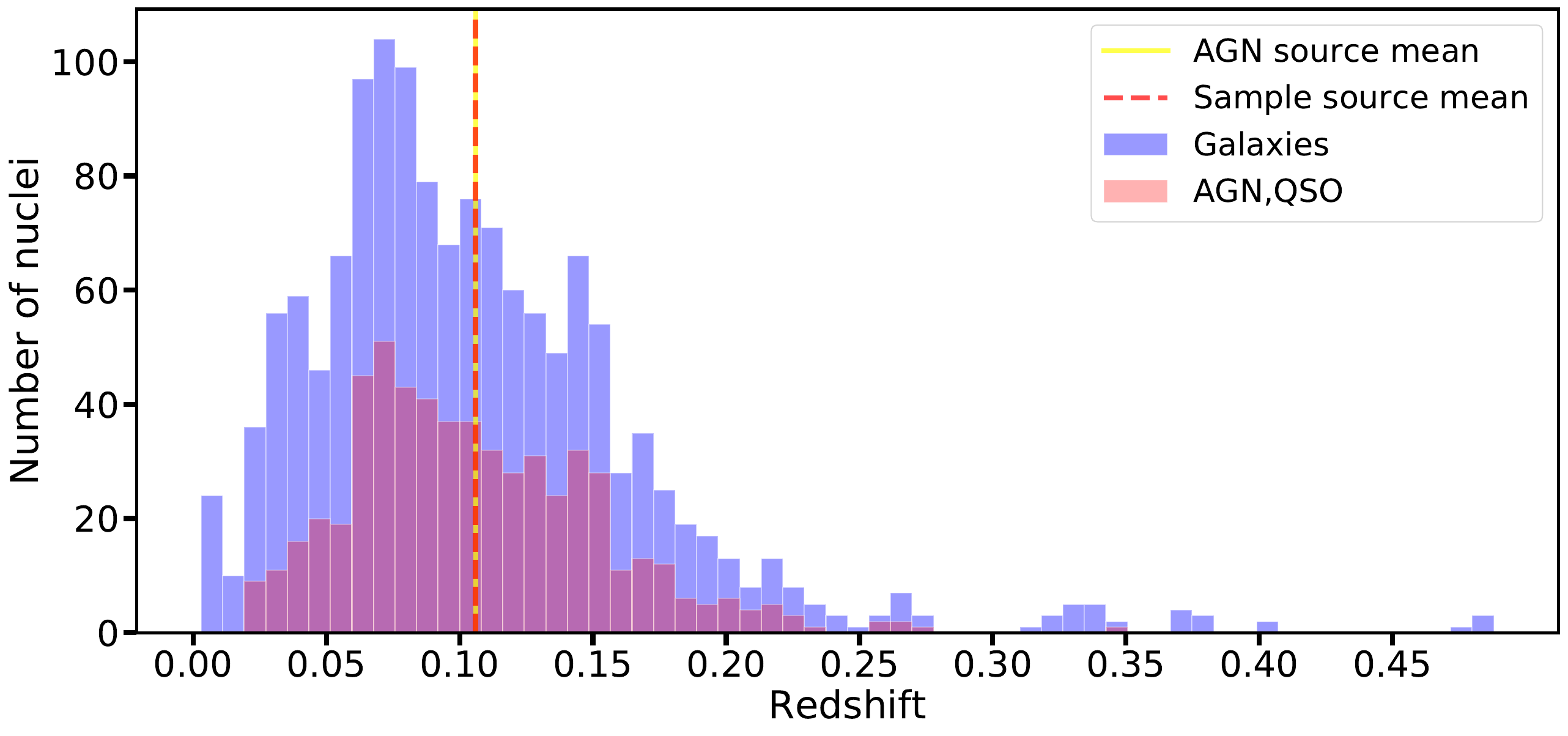}
        \vspace{-5pt}
        \caption{Confirmed DNG sample}
        \label{fig:redshift.confirmed}
    \end{subfigure}
    \caption{Redshift Distributions}
    \label{fig:redshift.histogram}
\end{figure*}

\subsection{Justification for Choosing a Spectroscopic Sample}
\label{sec:spectro.justice}



In our preliminary experiments of deploying \gothic to an unknown sample of galaxies, we found numerous detections of DNGs most of which turned out to be false positives. Upon manual investigation, these detections were essentially that of large foreground stars with diffraction spikes that dominated the space of the $40'' \times 40''$ cutout. We provide an example of such an object along with its DR16 link\footnote{
\url{https://skyserver.sdss.org/dr16/en/tools/explore/Summary.aspx?id=1237680332175769803}
}. According to SDSS, this object has its ``type'' attribute named as ``GALAXY'' with a valid photometric redshift. Using samples with a valid spectroscopic redshift would automatically eliminate such \textit{contaminants} in the starting blind sample. Moreover, the final outcome of running \gothic through a blind sample was to obtain a reduced sample which we could study for further insight. With this goal in mind, it was decided that the initial blind sample only contain galaxies with available spectroscopic redshift. Further justifications for choosing a spectroscopic \mybf{sample} have been provided in the Supplementary Section 1.5.1.

\begin{table}
\centering
\caption{Frequency of each possible combination}
\label{table:combin}
\begin{tabular}{|l|c|}
\hline
\textbf{combination}     &  \textbf{count} \\
\hline
 AGN-AGN    & 153\\
 \hline
 AGN-AGN-AGN & 2\\
 \hline
 AGN-AGN-SF-SF & 1\\
 \hline
 AGN-AGN-SF & 3\\
 \hline
 AGN-SF & 150\\
 \hline
 AGN-SF-SF & 1\\
 \hline
 SF-SF & 152 \\
 \hline
 SF-SF-SF & 2\\
 \hline
 \end{tabular}
\end{table}

\section{Analysis of the Reduced Candidate Sample}
\label{sec:analysis}

\subsection{The final sample of galaxy nuclei pairs}

We visually checked the sample of 949 nuclei pairs to determine the success rate of \gothic. We found that 681 were true galaxy nuclei pairs, which gives a success rate of $~71\%$. The false detection were mainly because of the following reasons.
\begin{enumerate}
    \item The target galaxy did not have a compact bulge or nucleus and so \gothic incorrectly classified it as a merger (e.g. \objid : \href{https://skyserver.sdss.org/dr12/en/tools/explore/summary.aspx?id=1237663239272923278}{1237663239272923278})
    \item The companion galaxy did not have a redshift, even though it was classified as a spectroscopic source (e.g. \objid : \href{https://skyserver.sdss.org/dr12/en/tools/explore/summary.aspx?id=1237671124306231474}{1237671124306231474})
    \item The system appears to be a merger but the second galaxy does not have a clear nucleus (e.g. \objid : \href{https://skyserver.sdss.org/dr12/en/tools/explore/summary.aspx?id=1237658204507340929}{1237658204507340929}). 
\end{enumerate}

\subsection{The redshift distribution of the initial and final samples}
\figureref{redshift.mill} shows the redshift distribution of the initial sample of one million galaxies from SDSS~DR16. The redshift extends from \(z=0\) to \(0.832\) and shows three peaks at approximately \(z=0.079\), \(0.353\), \(0.494\). The redshift distribution of the confirmed DNG (\figureref{redshift.confirmed}) is however slightly different and is distributed mainly between redshifts of \(z=0\) to \(0.3\). It shows only one peak at approximately \(z=0.073\), which is close to the first peak of the one million galaxy distribution, but does not have the exact same value. A significant fraction of the 681 sources are classified as DAGN pairs using SDSS and their distribution is overlaid in pink. The DAGN have the same distribution as the nuclei pairs. The mean redshifts of the nuclei pairs and DAGN are very close to each other and are at $z\approx$0.1 as shown by the dashed vertical line in \figureref{redshift.confirmed}.

\subsection{The classification of nuclei pairs using emission line ratios (BPT plot)}
As described in the introduction, our aim was to find a sample of DAGN  \citep{rubinur.mrk212.2020} or multiple AGN \citep{yadav.etal.2021} systems. To obtain such a sample, the SDSS class of the nuclei was very important. The SDSS classifies nuclei according to the emission lines in the spectra and the commonly used classes are starforming, starburst, AGN and QSO \mybf{\citep{Jonsson.etal.2020}.} If an AGN has broad components in the emission lines with doppler widths $>$200 kms$^{-1}$ it is classified as AGN broadline. The quasars (QSO), which are the most powerful AGN and have very broad emission lines, are identified using spectral templates. 

In our sample of 949 nuclei pairs, 681 were visually confirmed to be nuclei pairs.  We also found that many of the nuclei pairs in our sample are in small groups. This is not surprising as the clustering of galaxies has been known for decades \citep{dressler.1984}. Using the SDSS class, we found that 11 nuclei pairs were DAGN, and either QSO-QSO or AGN-QSO pairs. The subclass of the nuclei were extracted from SDSS wherever it was available and this gave finally 45 DAGN. But we found that SDSS DR16 had tabulated the class and subclass for relatively few sources and hence this gave us a low number of DAGN which did not seem to be complete. 

\begin{figure*}
        \centering
        \includegraphics[width=10cm,height=7.5cm]{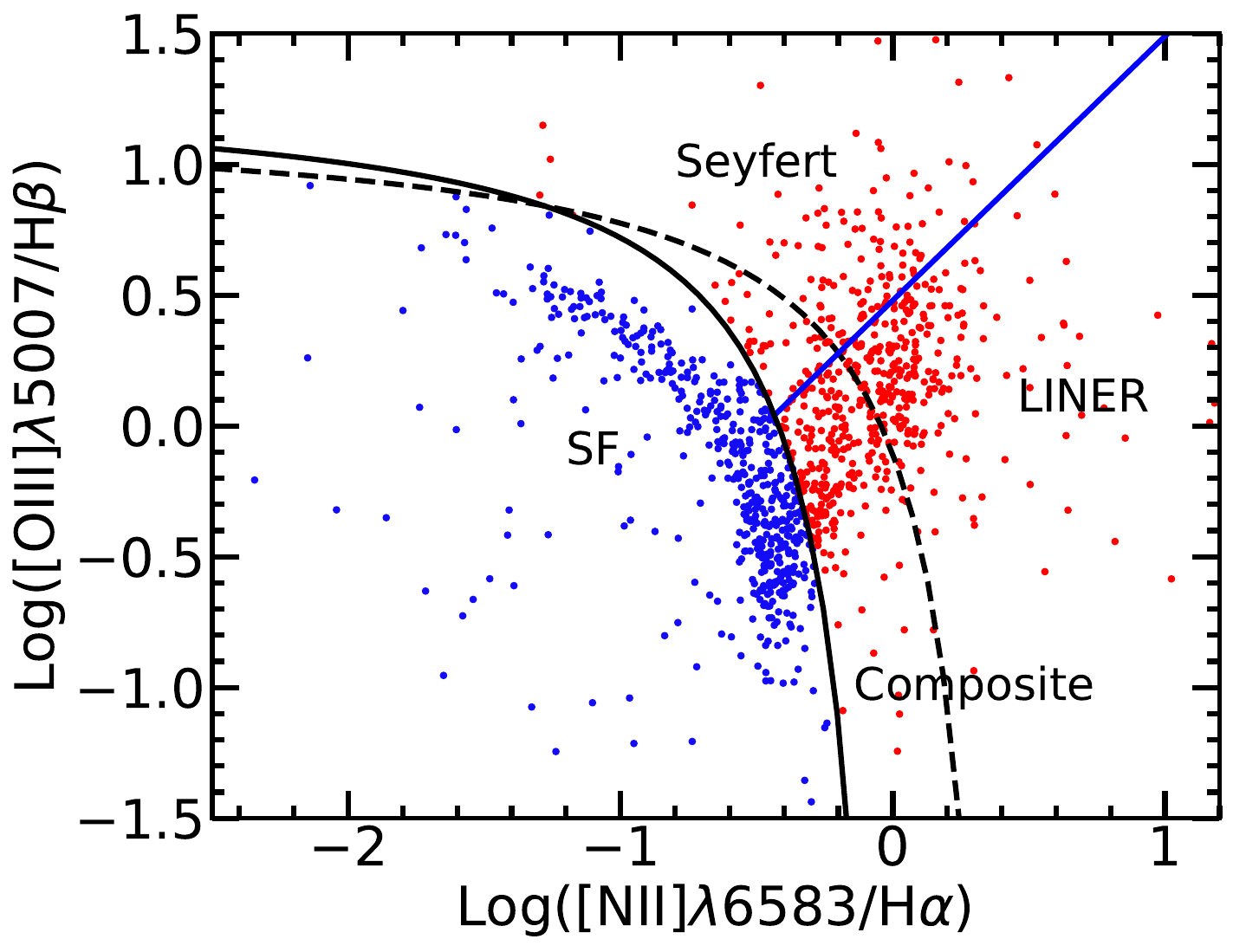}
        \caption{The BPT Diagram of the emission line data of 1098 nuclei that lie in pairs or small groups i.e. they are derived from the sample of 681 DNGs. From the position of the nuclei, their nature could be explained with reference to classification curves: The black solid line is the curve fit signifying the lower limit for AGNs \citep{kauffmann_host_2003}, hence the region below this curve is considered as SF (marked by blue dots). Similarly, the region above this curve is considered as AGN (marked by red dots). The black dashed curve is the upper limit for star burst \citep{kewley.etal.2006},sources in region between the solid black curve and dashed black curve are considered composite. The solid blue line separated LINER from seyferts \citep{cid.fernandes.etal.2010}} The distribution obtained is as follows : \mybf{SF=517, composite=244, LINER=232 and Seyfert=105}
\label{fig:bpt}
\end{figure*}
\begin{figure*}
    \centering
    \begin{subfigure}[b]{0.15\textwidth}
         \centering
         \includegraphics[width=\textwidth]{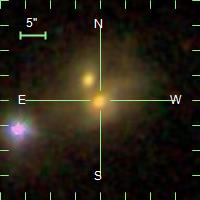}
         \caption*{1237650762394959890}
     \end{subfigure}%
     \hfill
     \begin{subfigure}[b]{0.15\textwidth}
         \centering
         \includegraphics[width=\textwidth]{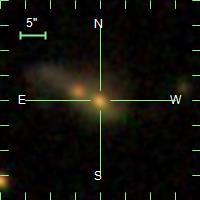}
         \caption*{1237651250964004981}
     \end{subfigure}%
     \hfill
     \begin{subfigure}[b]{0.15\textwidth}
         \centering
         \includegraphics[width=\textwidth]{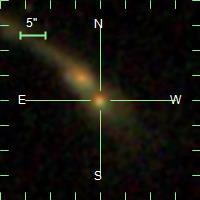}
         \caption*{1237655465920364786}
     \end{subfigure}%
     \hfill
     \begin{subfigure}[b]{0.15\textwidth}
         \centering
         \includegraphics[width=\textwidth]{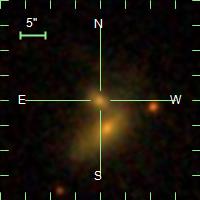}
         \caption*{1237655692474515648}
     \end{subfigure}%
     \hfill
     \begin{subfigure}[b]{0.15\textwidth}
         \centering
         \includegraphics[width=\textwidth]{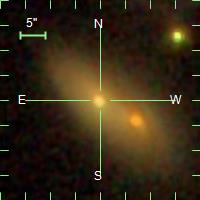}
         \caption*{1237656567585833094}
     \end{subfigure}%
     \hfill
     \begin{subfigure}[b]{0.15\textwidth}
         \centering
         \includegraphics[width=\textwidth]{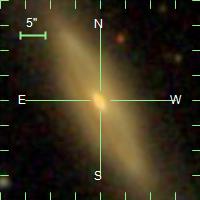}
         \caption*{1237656906354393114}
    \end{subfigure}
    
    \begin{subfigure}[b]{0.15\textwidth}
         \centering
         \includegraphics[width=\textwidth]{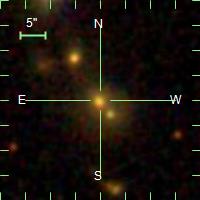}
         \caption*{1237657189817516083}
     \end{subfigure}%
     \hfill
     \begin{subfigure}[b]{0.15\textwidth}
         \centering
         \includegraphics[width=\textwidth]{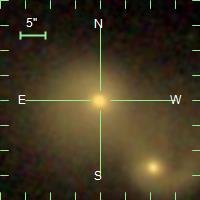}
         \caption*{1237657221485625353}
     \end{subfigure}%
     \hfill
     \begin{subfigure}[b]{0.15\textwidth}
         \centering
         \includegraphics[width=\textwidth]{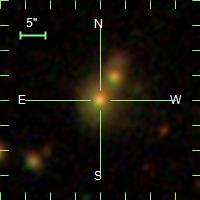}
         \caption*{1237657609104982532}
     \end{subfigure}%
     \hfill
     \begin{subfigure}[b]{0.15\textwidth}
         \centering
         \includegraphics[width=\textwidth]{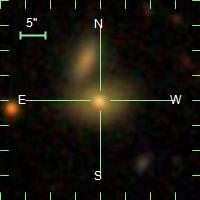}
         \caption*{1237657612874940605}
     \end{subfigure}%
     \hfill
     \begin{subfigure}[b]{0.15\textwidth}
         \centering
         \includegraphics[width=\textwidth]{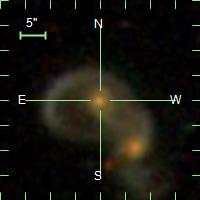}
         \caption*{1237658300056338796}
     \end{subfigure}%
     \hfill
     \begin{subfigure}[b]{0.15\textwidth}
         \centering
         \includegraphics[width=\textwidth]{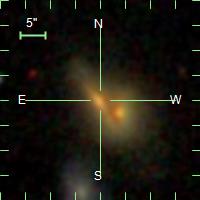}
         \caption*{1237658424090493080}
     \end{subfigure}

    \begin{subfigure}[b]{0.15\textwidth}
         \centering
         \includegraphics[width=\textwidth]{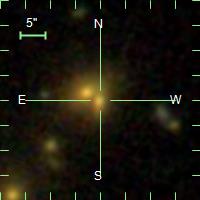}
         \caption*{1237658611440222365}
     \end{subfigure}%
     \hfill
     \begin{subfigure}[b]{0.15\textwidth}
         \centering
         \includegraphics[width=\textwidth]{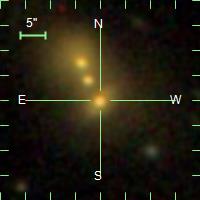}
         \caption*{1237661871330361452}
     \end{subfigure}%
     \hfill
     \begin{subfigure}[b]{0.15\textwidth}
         \centering
         \includegraphics[width=\textwidth]{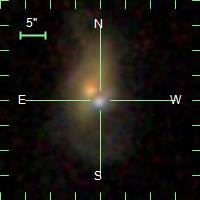}
         \caption*{1237665331455459379}
     \end{subfigure}%
     \hfill
     \begin{subfigure}[b]{0.15\textwidth}
         \centering
         \includegraphics[width=\textwidth]{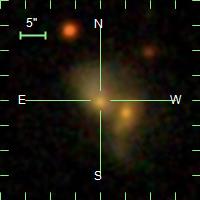}
         \caption*{1237659119862939753}
     \end{subfigure}%
     \hfill
     \begin{subfigure}[b]{0.15\textwidth}
         \centering
         \includegraphics[width=\textwidth]{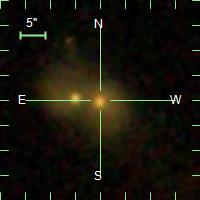}
         \caption*{1237661957762842806}
     \end{subfigure}%
     \hfill
     \begin{subfigure}[b]{0.15\textwidth}
         \centering
         \includegraphics[width=\textwidth]{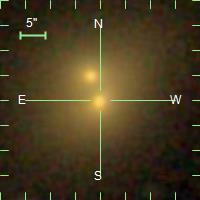}
         \caption*{1237665440439336990}
     \end{subfigure}

    \begin{subfigure}[b]{0.15\textwidth}
         \centering
         \includegraphics[width=\textwidth]{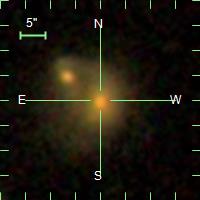}
         \caption*{1237659119864774675}
     \end{subfigure}%
     \hfill
     \begin{subfigure}[b]{0.15\textwidth}
         \centering
         \includegraphics[width=\textwidth]{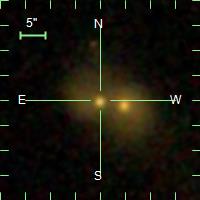}
         \caption*{1237661957762842807}
     \end{subfigure}%
     \hfill
     \begin{subfigure}[b]{0.15\textwidth}
         \centering
         \includegraphics[width=\textwidth]{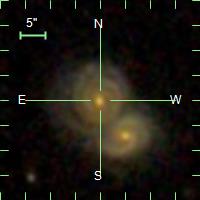}
         \caption*{1237667549268213959}
     \end{subfigure}%
     \hfill
     \begin{subfigure}[b]{0.15\textwidth}
         \centering
         \includegraphics[width=\textwidth]{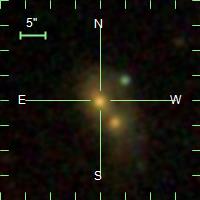}
         \caption*{1237659132206448648}
     \end{subfigure}%
     \hfill
     \begin{subfigure}[b]{0.15\textwidth}
         \centering
         \includegraphics[width=\textwidth]{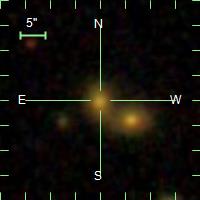}
         \caption*{1237662193453760670}
     \end{subfigure}%
     \hfill
     \begin{subfigure}[b]{0.15\textwidth}
         \centering
         \includegraphics[width=\textwidth]{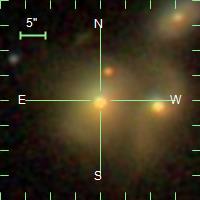}
         \caption*{1237667783929299179}
     \end{subfigure}

    \begin{subfigure}[b]{0.15\textwidth}
         \centering
         \includegraphics[width=\textwidth]{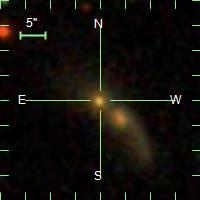}
         \caption*{1237660240313385134}
     \end{subfigure}%
     \hfill
     \begin{subfigure}[b]{0.15\textwidth}
         \centering
         \includegraphics[width=\textwidth]{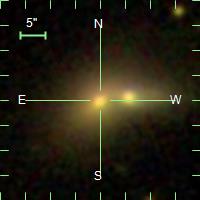}
         \caption*{1237662194541002843}
     \end{subfigure}%
     \hfill
     \begin{subfigure}[b]{0.15\textwidth}
         \centering
         \includegraphics[width=\textwidth]{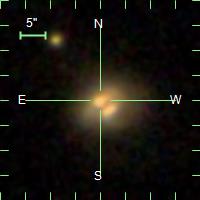}
         \caption*{1237667910069190795}
     \end{subfigure}%
     \hfill
     \begin{subfigure}[b]{0.15\textwidth}
         \centering
         \includegraphics[width=\textwidth]{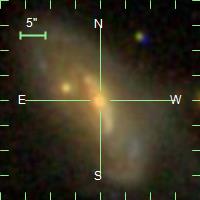}
         \caption*{1237660240313712760}
     \end{subfigure}%
     \hfill
     \begin{subfigure}[b]{0.15\textwidth}
         \centering
         \includegraphics[width=\textwidth]{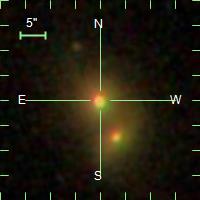}
         \caption*{1237662200436752504}
     \end{subfigure}%
     \hfill
     \begin{subfigure}[b]{0.15\textwidth}
         \centering
         \includegraphics[width=\textwidth]{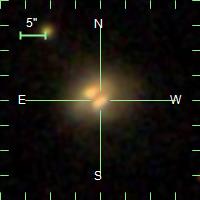}
         \caption*{1237667910069190796}
     \end{subfigure}

    \begin{subfigure}[b]{0.15\textwidth}
         \centering
         \includegraphics[width=\textwidth]{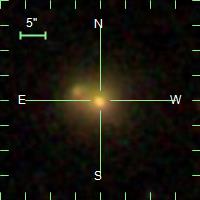}
         \caption*{1237660241385029806}
     \end{subfigure}%
     \hfill
     \begin{subfigure}[b]{0.15\textwidth}
         \centering
         \includegraphics[width=\textwidth]{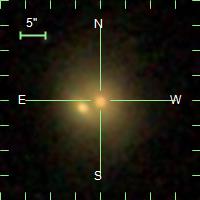}
         \caption*{1237662336792920257}
     \end{subfigure}%
     \hfill
     \begin{subfigure}[b]{0.15\textwidth}
         \centering
         \includegraphics[width=\textwidth]{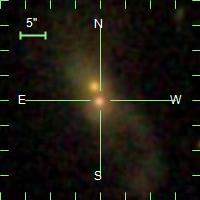}
         \caption*{1237668298204446938}
     \end{subfigure}%
     \hfill
     \begin{subfigure}[b]{0.15\textwidth}
         \centering
         \includegraphics[width=\textwidth]{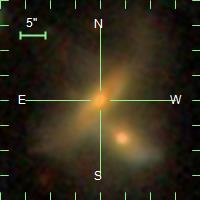}
         \caption*{1237660963472932915}
     \end{subfigure}%
     \hfill
     \begin{subfigure}[b]{0.15\textwidth}
         \centering
         \includegraphics[width=\textwidth]{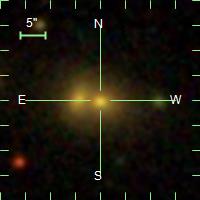}
         \caption*{1237662498390737019}
     \end{subfigure}%
     \hfill
     \begin{subfigure}[b]{0.15\textwidth}
         \centering
         \includegraphics[width=\textwidth]{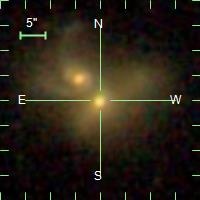}
         \caption*{1237668588651282620}
     \end{subfigure}

    \begin{subfigure}[b]{0.15\textwidth}
         \centering
         \includegraphics[width=\textwidth]{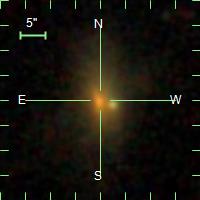}
         \caption*{1237661151377424586}
     \end{subfigure}%
     \hfill
     \begin{subfigure}[b]{0.15\textwidth}
         \centering
         \includegraphics[width=\textwidth]{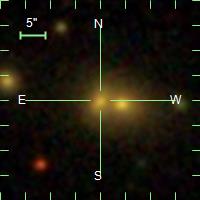}
         \caption*{1237662498390737020}
     \end{subfigure}%
     \hfill
     \begin{subfigure}[b]{0.15\textwidth}
         \centering
         \includegraphics[width=\textwidth]{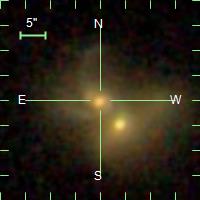}
         \caption*{1237668588651282621}
     \end{subfigure}%
     \hfill
     \begin{subfigure}[b]{0.15\textwidth}
         \centering
         \includegraphics[width=\textwidth]{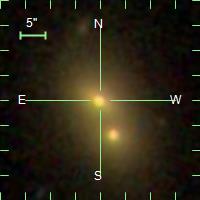}
         \caption*{1237661382778093715}
     \end{subfigure}%
     \hfill
     \begin{subfigure}[b]{0.15\textwidth}
         \centering
         \includegraphics[width=\textwidth]{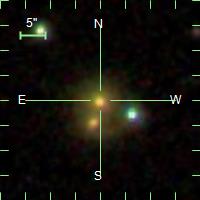}
         \caption*{1237663783136985125}
     \end{subfigure}%
     \hfill
     \begin{subfigure}[b]{0.15\textwidth}
         \centering
         \includegraphics[width=\textwidth]{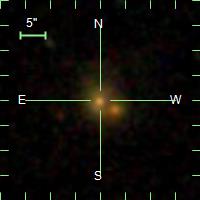}
         \caption*{1237668623551103255}
     \end{subfigure}
     \caption{\mybf{Panel of a subset of 42 out of the 159 discovered Dual AGNs (with SDSS \objids listed)}}
     \label{fig:panel}
\end{figure*}

So for a better classification of the nuclei, we instead used the SDSS tabulated emission line data of the nuclei to plot the Baldwin, Phillips and Terlevich (BPT) diagram of our 681 DNG sources \mybf{\citep{BPT.1981}}. The BPT classifies nuclei based on the ratio of emission lines such as H$\alpha$, [OIII], [SII], [NII]. Out of the total sample of 681 sources we have tried to incorporate the emission line data of all the nuclei for which data was available and obtained line data for 1393 nuclei (\figureref{bpt}). The classes of the nuclei are star forming (SF hereafter), composite, LINER and Seyfert. We have taken the solid curve in the \mybf{Figure 4} as the lower limit for finding an AGN and hence the classes composite, LINER and Seyfert are all considered to be AGN \citep{kauffmann_host_2003}. Studies have shown that the composite class does host AGN \citep{juneau.etal.2014}. 

However, there were some problems in obtaining the exact nature of some nuclei mainly because : 
\begin{enumerate}
    \item Out of 1393 nuclei 14 had unavailable line emission data
    \item 258 nuclei had either one of the emission line fluxes tabulated as negative or the denominator in the flux ratio was 0.
    \item 22 had negative emission line line fluxes but the ratio was positive
    \item One source had unusual [OIII] emission line values
\end{enumerate}
So finally we could classify the nature of 1098 nuclei, of which 581 were AGN in nature and 517 were star forming. So the final BPT plot had 1098 nuclei (\figureref{bpt}).     
We examined the nature of the sources in detail and found that 159 sources had at least two of the nuclei classified as AGN, so the number of DAGN detected in our sample is 159 (\tableref{combin}). However, it must be noted that this is a lower limit, as the spectral information for several nuclei sources was not available. Of these 159 DAGN, 2 are in clusters of triple AGN lying with groups of 3 to 4 galaxies. Triple AGN are rare and only a few have been detected using X-ray and optical observations \citep{yadav.etal.2021}. One DAGN is in a group of 4 nuclei with the other two being star forming in nature. One DAGN is in a triplet system where the third is a star forming nucleus. 

The  frequency of the different combinations obtained is given in \tableref{combin}. It is clear that the number of AGN-SF and DAGN are comparable. This is surprising as the number of star forming nuclei is generally found to be larger than the number of DAGN in previous studies (e.g. \citet{rubinur.etal.2019}). But our sample is composed of relatively close nuclei pairs, hence this suggests that star formation maybe quenched in galaxies as the nuclei come closer during mergers. \tableref{combin} also shows that the number of composite class nuclei is relatively large. These nuclei have star formation associated with AGN activity and may represent galaxies evolving from from starburst dominated galaxies to AGN dominated ones \citep{yuan.etal.2010}. The number of star forming (SF-SF) or AGN-SF pairs is also very large, implying that star formation is an integral part of the interaction process, but may not be the dominant one as the number of SF-SF nuclei and DAGN are comparable.

The full table of the 159 sources is available in \tableref{confirmed.dual.agns}. Also, a panel depicting a subset of the sources (42 of them) are presented in \figureref{panel}. 

\begin{figure*}
        \centering
        \includegraphics[width=12cm,height=11cm]{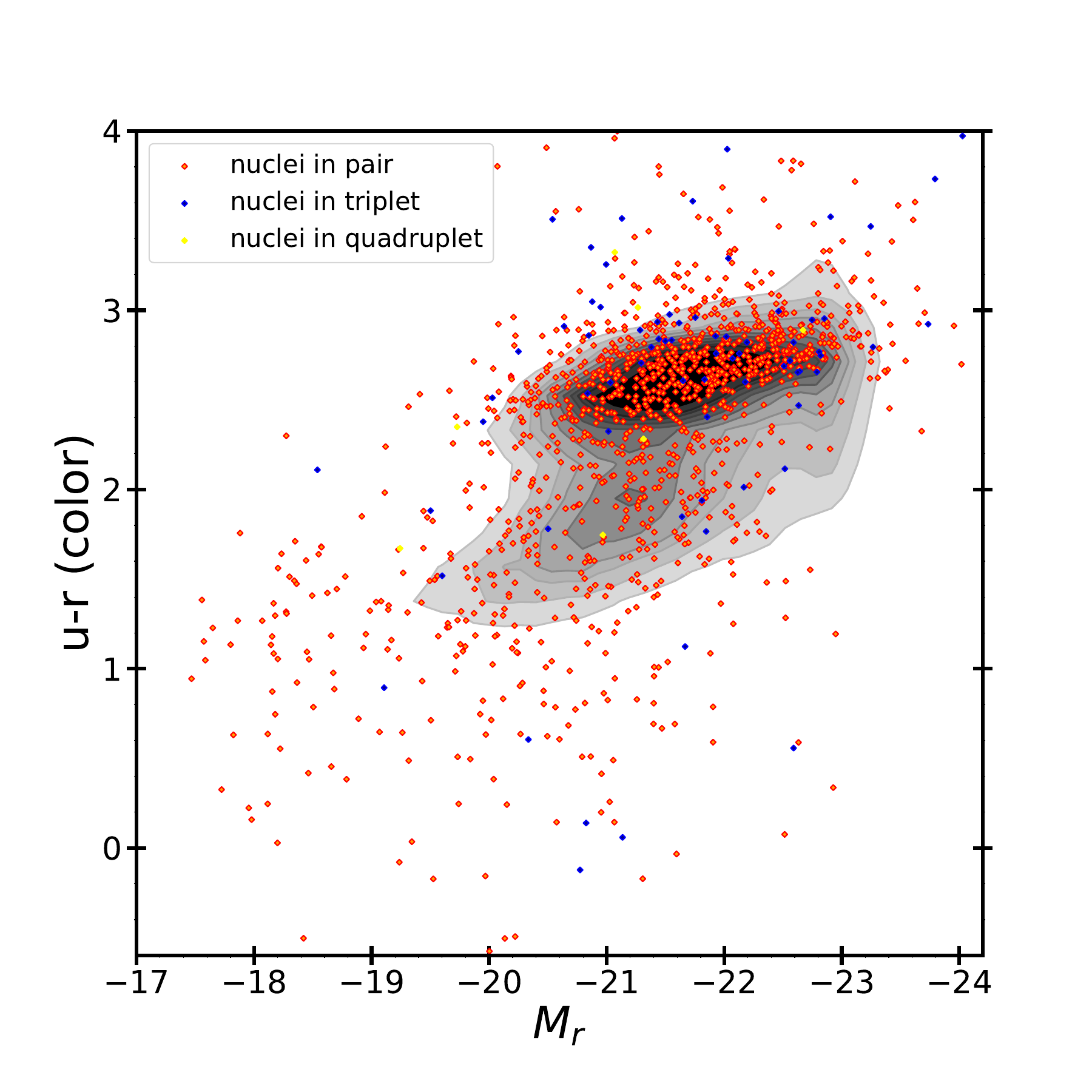}
        
        \caption{The color magnitude diagram (CMD) of all 1393 galaxies in pairs of small groups (i.e. the total number of sources from the 681 galaxies in pairs  and including their companions; a few are in groups of 3 or 4). \mybf{A comparison sample of 743,379 SDSS galaxies (with clean photometric data available \ie having a value of 1 in the \texttt{clean} attribute of the galaxy's corresponding entry in the \texttt{PhotoObjAll} table) at redshifts $z < 0.5$ are plotted in grey scale}. The integrated ($u − r$) colors corrected for extinction of the galaxies are plotted against their absolute magnitudes in the r band, \mybf{$M_r$}. Galaxies belonging to group having different number of companions are given different color (red for galaxies in group having 2 galaxies , blue for galaxies in group having 3 galaxies, yellow for galaxies in group having 4 galaxies) and are shown as individual colored points superimposed on a grey-scale contour map of the distribution of the comparison sample , where the darker shades represent the denser regions. The density contour levels are in units of the number of galaxies per grid unit of size 0.132 mag in \mybf{$M_r$} and 0.19 mag in ($u - r$): 1500, 2250, 3000, 3500, 4200, 4800, 5400, 6000, 6600, 7000, 8000, 10000.}
\label{fig:contour}
\end{figure*}
\begin{figure*}
\hspace{-0.25cm}
     \centering
     \begin{subfigure}[b]{0.33\textwidth}
         \centering
         \includegraphics[width=\textwidth]{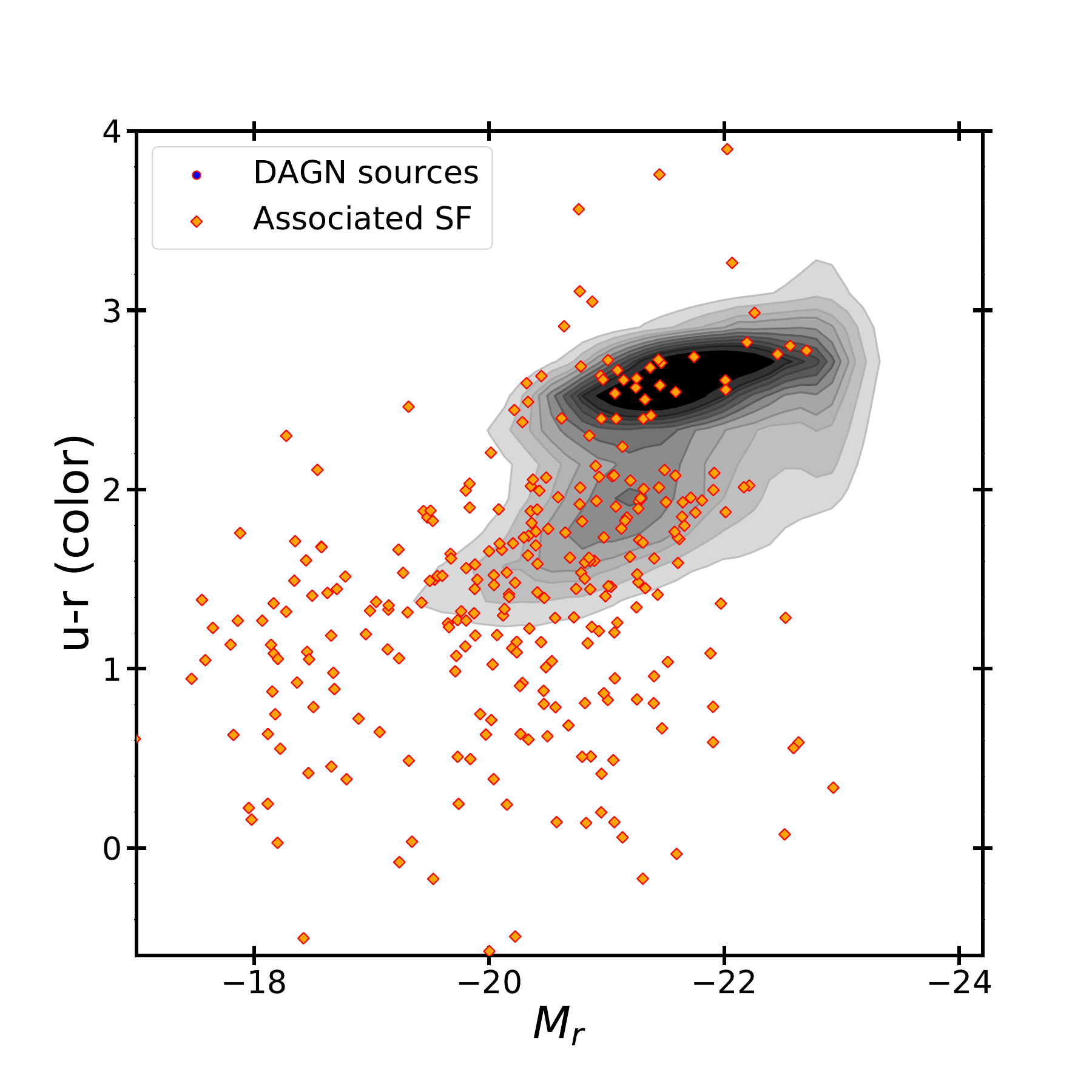}
         \caption{CMD for SF-SF pairs.}
         \label{fig:SF-SFpairs}
     \end{subfigure}
\hspace{-0.1cm}     
     \begin{subfigure}[b]{0.33\textwidth}
         \centering
         \includegraphics[width=\textwidth]{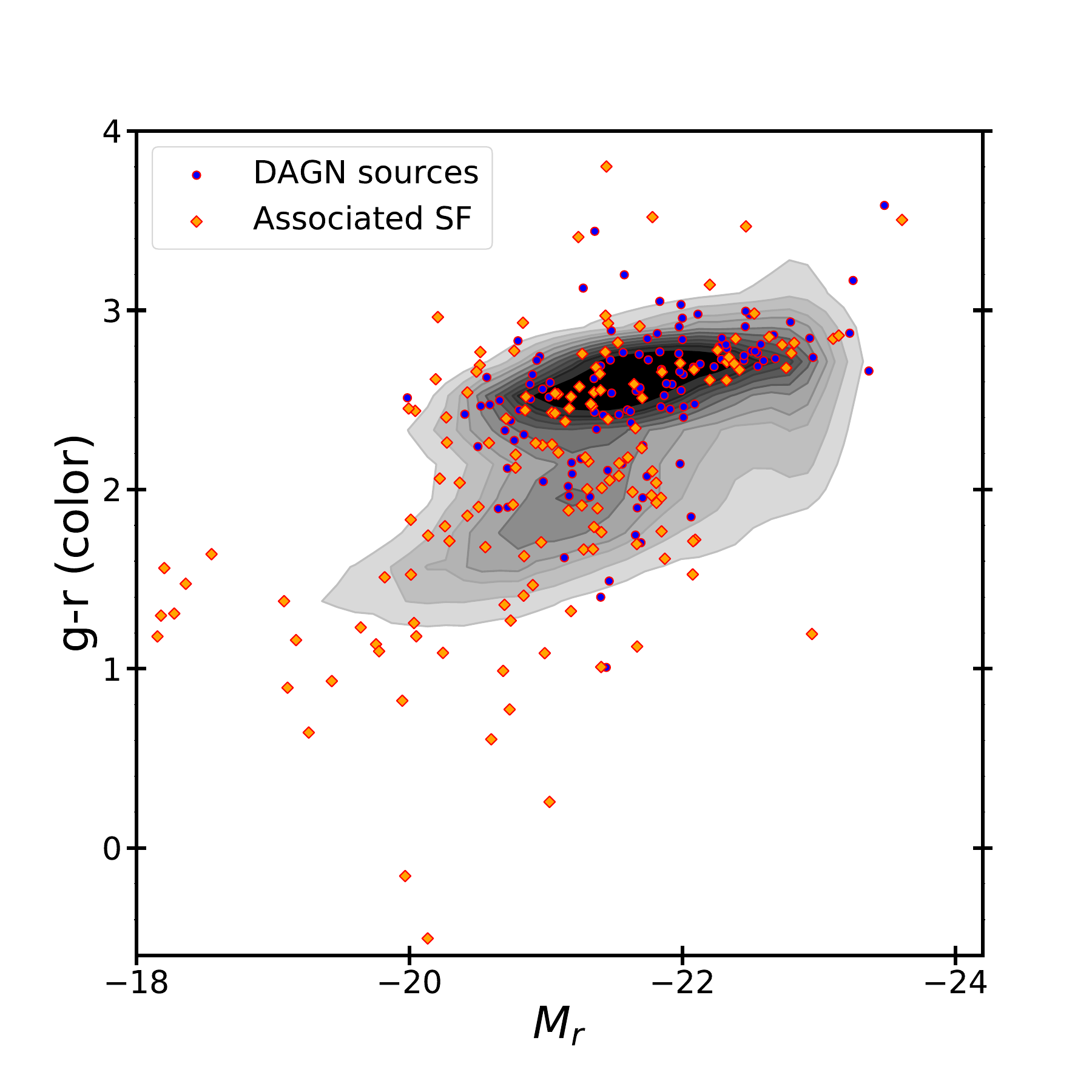}
         \caption{CMD for mixed pairs ie AGN-SF.}
         \label{fig:AGN-SFpairs}
     \end{subfigure}
\hspace{-0.1cm}      
     \begin{subfigure}[b]{0.33\textwidth}
         \centering
         \includegraphics[width=\textwidth]{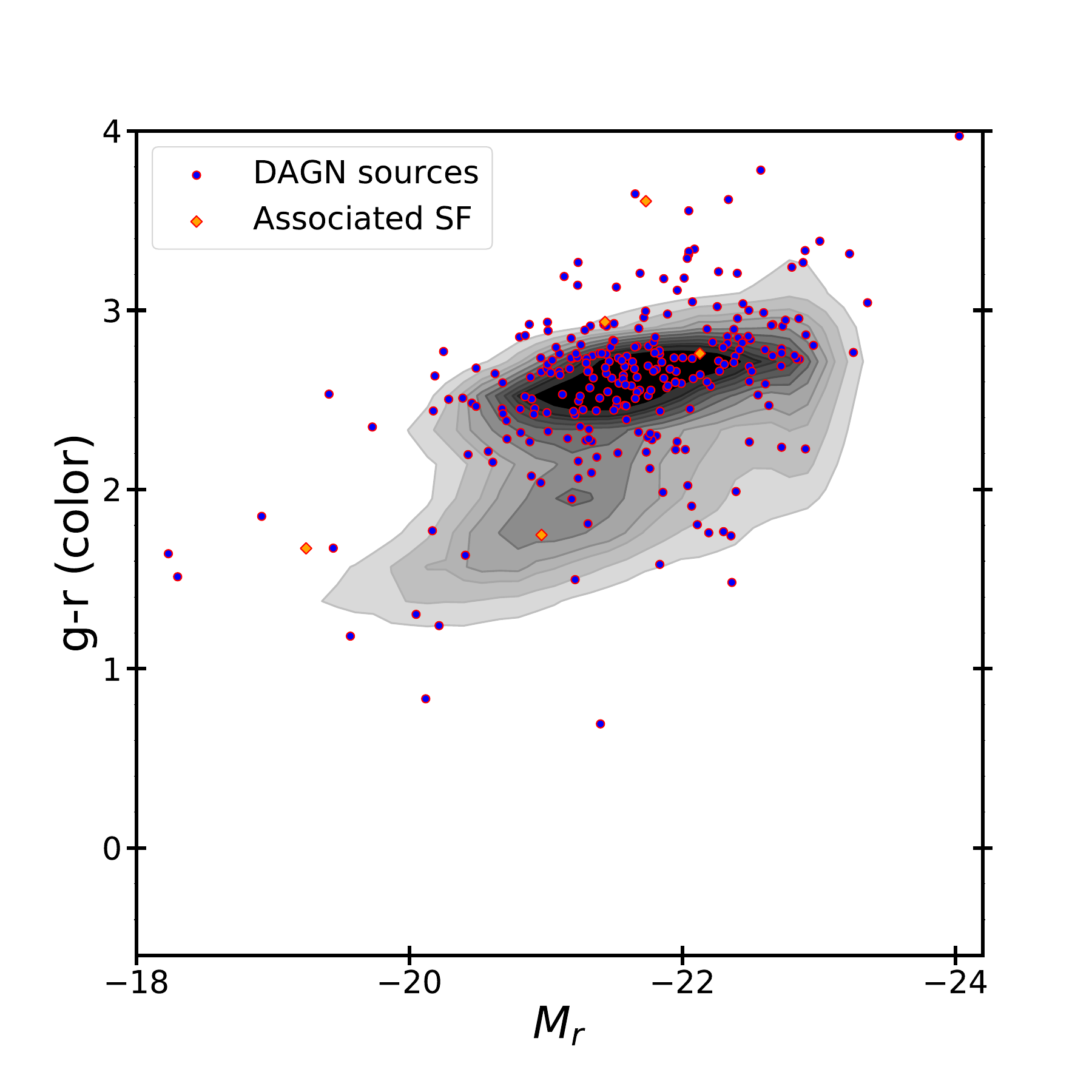}
         \caption{CMD for DAGN ie AGN-AGN.}
         \label{fig:AGNpairs}
     \end{subfigure}     
     
     \caption{The 3 plots from left to right show respectively the relative positions of the star forming nuclei pairs (SF-SF), the mixed AGN-SF pairs and the DAGN on the CMD. The DAGN includes the two triple AGN.}
\label{fig:allCMD}     
\end{figure*}

\subsection{The color magnitude plot of the galaxy nuclei pairs and comparison with the general population of SDSS galaxies}
The color magnitude diagram (CMD) of galaxies is usually a plot of the color of galaxies (e.g. $u - r$ magnitudes) against their absolute r band magnitude (\mybf{$M_r$}) and is known to show a bimodal distribution \citep{bell.etal.2004}. The ($u - r$) color represents the star formation rate (SFR) per unit stellar mass whereas \mybf{$M_r$} represents the stellar mass of a galaxy. The CMD is bimodal with the star forming galaxies lying in the bluer and hence lower part of the CMD whereas the redder galaxies with low SFRs are in the upper part of the CMD \citep{schawinski.etal.2014}. These two regions are traditionally called the blue cloud and red sequence respectively, although other wavebands colors such as ($NUV - r$) have also been used \citep{salim.etal.2007}. 

\figureref{contour} shows the CMD of all the galaxies from the 681 galaxies, including galaxies in pairs of small groups. The total number is 1393. A comparison sample of 743379 SDSS galaxies with clean photometry flag at redshifts \(z < 0.499\) are plotted in grey scale. The integrated ($u - r$) colors of the galaxies are plotted against their absolute magnitudes in the r band \mybf{$M_r$} after they were corrected for Galactic extinction. Galaxies belonging to group having different number of companions are given different color (red for galaxies in group having 2 galaxies, blue for galaxies in groups of 3 galaxies, black for groups of 4 galaxies) and are shown as individual colored points superimposed on a grey-scale contour map. It is interesting to see that a large fraction of the DNGs lie in the redder part of the CMD. This suggests that the star formation associated with the merging process is quenched as the nuclei come closer. The galaxies also have absolute magnitudes \mybf{$M_r<18$} which suggests that the DNGs are associated with relatively more massive galaxies. 


\subsection{Estimating the fraction of DAGN in the 1 million sample}

As shown in Section 5, applying \gothic to a sample of a million galaxies and the subsequent visual inspection results in a clean sample of 159 DAGN, all of which have spectroscopic confirmation. However, this is a lower limit and there are several ways in which ``true'' DAGN could have been missed in this process (e.g. spectroscopic data are not available, the components are closer than the minimum separation allowed by SDSS fibres, redshifts do not indicate pairs, noisy data). So it will be useful to have a ball park estimate of the number of DAGN expected in a sample of a million galaxies. This number will also tell us what is the number of DAGN we have missed in this study.

The number is difficult to estimate because many nuclei pairs can simply be due to the chance superposition of galaxies in the sky. But we can make a rough estimate of the upper limit to the expected number of DAGN based on the numbers we have obtained (see \figureref{filtration.flowchart}).
\mybf{ 
\begin{enumerate}
    \item The total number of nuclei pairs that \gothic determines from the million galaxy sample is 95,159. The number with redshifts is 47,521 (\(\ie 49.93\%\)), and from this sample 46,061 (\(96.92\%\)) are close pairs within a fibre and 949 (\ie \(1.99\%\)) have separate redshifts. But the upper limit to the number of nuclei pairs that could be true close pairs is 95,159.
    \item Since we need to estimate the number of expected DAGN we can use the fact that from the sample of 681 nuclei pairs we have detected 159 DAGN (as explained in Section 5). Hence the number of DAGN that could be expected from a one million sample \(= (159/681)\times 95,159 \approx 22,218\) DAGN (\ie \(2.2\%\)).
    \item Since we have detected 159 DAGN from the sample, an upper limit to the fraction that we have missed is (22,218-159)/22,218 \(\approx\) 99.3\% approximately.
\end{enumerate}
}
So DAGN are not only rare (2.2\% predicted detection rate), but also difficult to detect. Mainly because many nuclei pairs could be due to superposition, large fraction of nuclei pairs lie within the SDSS fibre and a significant fraction of nuclei pairs do not have spectroscopic redshifts.

\section{Discussion}
\label{sec:discussion}
In this study we have discovered a new sample of dual/triple AGN, apart from mixed AGN-star forming nuclei pairs. We have used a novel algorithm named \gothic to automatically create a pool of DNGs and confirmed their nature using empirical relations such as those used in the BPT plot. Such algorithms are important for finding SMBH pairs in large data sets \citep{manucci.eta.2022}. The final list of DAGN we obtained could be important inputs for models that predict the gravitation wave background from SMBH binaries \citep{casey-clyde.etal.2022}.

\mybf{As discussed in the previous section, the number of DAGN that we expect (2.2\%) in any galaxy population is much smaller than predicted in simulation studies \citep{volonteri.etal.2022,bhowmick.etal.2020}.} 
Since the redshift distribution extends from \(z=0\) to \(0.7\), this translates approximately to a typical projected separation of a few kpc at \(z=0\) to a maximum of $10^{\prime\prime}$ or 145kpc at \(z=0.7\) (where $H_{0}=70kms^{-1}Mpc^{-1}$). The number of DAGN is 159, of which two are triple AGN. These numbers however, are lower limits, since we had to exclude nuclei pairs that lie within the SDSS fibre radius of $1.5^{\prime\prime}$ as well as those that \mybf{did not} have good \mybf{signal-to-noise ratio} (S/N) spectroscopic data and so reliable emission line ratios could not be obtained. But overall these numbers suggest that dual/multiple AGN  systems are not common (\tableref{combin}). The DAGN reported in the literature \citep{das.etal.2018} are often serendipitous in nature, and multi-wavelength surveys usually do not yield many DAGN \citep{rubinur.etal.2019}. But AGN are known to produce winds/outflows that are important for building up the hot gas halos or CGM around galaxies and the presence of DAGN would increase this effect. Since the numbers of DAGN detected are relatively low, our results suggest that the starburst winds/outflows in galaxy mergers, and not just the DAGN, maybe be important for the growth of CGM around galaxies.

\mybf{Another interesting implication of our study is that a large fraction of the DNGs and most of the DAGNs lie in the red cloud region of the galaxy color magnitude plot (\figureref{contour}). To examine this in more detail we plotted the galaxy nuclei according to their type. \figureref{allCMD} shows the color magnitude diagram (CMD) for the different double nuclei pairs, star forming (SF-SF), mixed (AGN-SF) and DAGN. The SF-SF pairs lie over a wide range of colors (\(g - r\)), with a significant fraction lying in the blue region of the CMD. Whereas the DAGN are mainly found in the red region of the CMD. So as one moves from SF-SF pairs to AGN-SF pairs and then finally to DAGN, the colors become redder with the DAGN nuclei lying in the reddest part of the CMD. This suggests that as the nuclei become redder  during mergers (and perhaps come closer), star formation is quenched and AGN activity begins. Or perhaps AGN activity in DAGN quenches star formation. These effects have been observed in ultraluminous galaxies (ULIRGs) that are the result of gas rich major mergers and appear to be evolving into elliptical galaxies \citep{dasyra.etal.2006}. In some cases their nuclei may give out radio jets and evolve into radio galaxies as well \citep{nandi.etal.2021}.
}

The DAGN \mybf{host galaxies} are also more massive systems compared to the SF-SF and AGN-SF pair host galaxies. This indicates that the formation of the DAGN is the final stage in evolution of galaxy mergers and their nuclei. However, to understand this in more detail, the parameters of the nuclei, their separation and SMBH masses need to be determined. It will also be interesting to see how the DAGN evolve in cosmological simulations and if they show similar trends in the CMD. These questions will be followed up in our subsequent paper. Recall that \mybf{ $~46,061$ DNG detections} were rejected in the third filtration process (\sectionref{filtration}), however, this rejected sample could also potentially contain DAGNs and could be studied further.  

\section{Conclusion}
\label{sec:conclusion}

We present a novel detection algorithm called \gothic, that detects DNGs in given galaxy images from SDSS DR16. From this sample of DNGs, our goal was to find DAGN systems. The algorithm has been tested on an existing catalog galaxy pairs that host double nuclei, and the results show a $100\%$ detection rate. Subsequently, we applied \gothic on an \emph{unknown} sample of one million galaxies that have  spectroscopic data in SDSS DR16. From this sample, \mybf{nearly $100,000$} galaxies were reported to be DNG by \gothic. After careful filtration of the detected sample and removal of likely false detections, a small subset of 949 galaxies was studied visually to produce a confirmed sample of DAGN systems. We detected 681 nuclei pairs, and then used their emission line fluxes to determine their positions (AGN or star forming) on the BPT plot. The BPT showed that 159 are DAGN, of which 2 are triple AGN systems. The others are either star forming pairs (SF-SF) or AGN-star forming (AGN-SF) pairs. The color magnitude plot of the different nuclei pairs shows that the DAGN usually lie in galaxies that are redder and more massive compared to the SF-SF and AGN-SF pairs. Our results also suggest that DAGN may play a role in quenching star formation in galaxy mergers. 

\section*{Acknowledgements}

The first author gratefully acknowledges the support received from the Indian Institute of Astrophysics, Bangalore, India during an internship. The second author also acknowledges the support of the Indian Institute of Astrophysics, Bangalore, India for a summer internship. M.Das  acknowledges the support of the Science and Engineering Research Board (SERB) MATRICS grant MTR/2020/000266 for this research. S.Saha would like to thank the Science and Engineering research Board (SERB), Department of Science and
Technology, Government of India, for supporting our research by providing us with resources to conduct our study. The project reference number is EMR/2016/005687.

\bigskip

Funding for the SDSS and SDSS-II has been provided by the Alfred P. Sloan Foundation, the Participating Institutions, the National Science Foundation, the U.S. Department of Energy, the National Aeronautics and Space Administration, the Japanese Monbukagakusho, the Max Planck Society, and the Higher Education Funding Council for England. The SDSS Web Site is http://www.sdss.org/. The SDSS is managed by the Astrophysical Research Consortium for the Participating Institutions. The Participating Institutions are the American Museum of Natural History, Astrophysical Institute Potsdam, University of Basel, University of Cambridge, Case Western Reserve University, University of Chicago, Drexel University, Fermilab, the Institute for Advanced Study, the Japan Participation Group, Johns Hopkins University, the Joint Institute for Nuclear Astrophysics, the Kavli Institute for Particle Astrophysics and Cosmology, the Korean Scientist Group, the Chinese Academy of Sciences (LAMOST), Los Alamos National Laboratory, the Max-Planck- Institute for Astronomy (MPIA), the Max-Planck-Institute for Astrophysics (MPA), New Mexico State University, Ohio State University, University of Pittsburgh, University of Portsmouth, Princeton University, the United States Naval Observatory, and the University of Washington.

\section*{Data Availability}

All the data upon which this is work is based can be found at \\
\url{https://github.com/anuwu/Blindtest-HQ}.


\clearpage
\onecolumn
\begin{longtable}{|ccccccccc|}
\label{table:confirmed.dual.agns}\\
\hline
Nuclei ID & objID & \begin{tabular}[c]{@{}c@{}}Spectroscopic\\  redshift (z)\end{tabular} & u & r & \begin{tabular}[c]{@{}c@{}}\mybf{Absolute r-band}\\  \mybf{magnitude}\end{tabular} & class & \begin{tabular}[c]{@{}c@{}}subclass \\ (SDSS)\end{tabular} & \begin{tabular}[c]{@{}c@{}}subclass \\ (BPT)\end{tabular} \\
\hline
\endhead
GTC\_1\_1 & 1237650762394959890 & 0.081 & 18.63 & 15.65 & -22.31 & GALAXY & BROADLINE & Seyfert \\
GTC\_1\_2 & 1237650762394959891 & 0.081 & 19.58 & 16.83 & -21.12 & GALAXY & ... & Composite \\
\hline
GTC\_2\_1 & 1237651250964004981 & 0.102 & 20.31 & 17.51 & -21.88 & GALAXY & STARFORMING & Composite \\
GTC\_2\_2 & 1237651250964004982 & 0.102 & 19.82 & 17.48 & -21.23 & GALAXY & STARBURST & Composite \\
\hline
GTC\_3\_1 & 1237651252018151484 & 0.075 & 18.87 & 15.53 & -22.44 & GALAXY & ... & Composite \\
GTC\_3\_2 & 1237651252018151487 & 0.075 & 19.52 & 16.17 & -22.07 & GALAXY & ... & LINER \\
\hline
GTC\_4\_1 & 1237651252024377492 & 0.089 & 19.6 & 16.62 & -21.59 & GALAXY & ... & LINER \\
GTC\_4\_2 & 1237651252024377493 & 0.089 & 19.8 & 16.8 & -21.44 & GALAXY & ... & LINER \\
\hline
GTC\_5\_1 & 1237651273510289469 & 0.118 & 18.85 & 15.75 & -22.95 & GALAXY & ... & LINER \\
GTC\_5\_2 & 1237651273510289470 & 0.119 & 20.64 & 17.93 & -21.51 & GALAXY & ... & Composite \\
\hline
GTC\_6\_1 & 1237652600110383327 & 0.166 & 20.61 & 17.91 & -20.81 & GALAXY & STARFORMING & Composite \\
GTC\_6\_2 & 1237652600110383328 & 0.166 & 20.52 & 17.75 & -20.90 & GALAXY & ... & Seyfert \\\hline
GTC\_7\_1 & 1237652900234133595 & 0.200 & 21.76 & 17.74 & -22.80 & GALAXY & ... & LINER \\
GTC\_7\_2 & 1237652900234133596 & 0.198 & 22.36 & 19.1 & -20.18 & GALAXY & STARFORMING & Composite \\\hline
GTC\_8\_1 & 1237652946379604015 & 0.166 & 20.62 & 17.77 & -22.05 & GALAXY & ... & Composite \\
GTC\_8\_2 & 1237652946379604014 & 0.164 & 20.74 & 16.83 & -22.89 & GALAXY & ... & LINER \\\hline
GTC\_9\_1 & 1237653441374453924 & 0.069 & 19.29 & 16.3 & -21.34 & GALAXY & STARFORMING & Composite \\
GTC\_9\_2 & 1237653441374453925 & 0.071 & 18.58 & 15.71 & -21.56 & GALAXY & ... & Composite \\\hline
GTC\_10\_1 & 1237654031400566817 & 0.040 & 18.01 & 15.05 & -21.53 & GALAXY & ... & LINER \\
GTC\_10\_2 & 1237654031400566818 & 0.040 & 17.78 & 15.32 & -22.48 & GALAXY & STARFORMING & Composite \\\hline
GTC\_11\_1 & 1237654382516240489 & 0.125 & 19.28 & 16.16 & -23.25 & GALAXY & ... & LINER \\
GTC\_11\_2 & 1237654382516240490 & 0.123 & 20.23 & 16.98 & -21.42 & GALAXY & ... & Seyfert \\\hline
GTC\_12\_1 & 1237654880736968887 & 0.098 & 20.23 & 16.98 & -21.01 & GALAXY & ... & Composite \\
GTC\_12\_2 & 1237654880736968888 & 0.098 & 19.29 & 17.33 & -21.30 & GALAXY & STARFORMING & Composite \\\hline
GTC\_13\_1 & 1237655106765521018 & 0.114 & 19.39 & 16.26 & -22.46 & GALAXY & ... & Composite \\
GTC\_13\_2 & 1237655106765521019 & 0.113 & 20.09 & 17.82 & -21.33 & GALAXY & ... & Composite \\\hline
GTC\_14\_1 & 1237655369279733929 & 0.073 & 19.67 & 16.98 & -20.89 & GALAXY & STARFORMING & Composite \\
GTC\_14\_2 & 1237655369279733928 & 0.074 & 18.5 & 16.07 & -21.29 & GALAXY & STARFORMING & Composite \\\hline
GTC\_15\_1 & 1237655465384214778 & 0.074 & 18.04 & 15.07 & -22.66 & GALAXY & ... & Composite \\
GTC\_15\_2 & 1237655465384214780 & 0.071 & 19.94 & 17.19 & -21.68 & GALAXY & ... & Composite \\\hline
GTC\_16\_1 & 1237655465920364786 & 0.098 & 20.15 & 17.06 & -21.48 & GALAXY & ... & Seyfert \\
GTC\_16\_2 & 1237655465920364787 & 0.098 & 20.7 & 18.04 & -20.67 & GALAXY & ... & Composite \\\hline
GTC\_17\_1 & 1237655495979106559 & 0.091 & 20.11 & 17.34 & -20.84 & GALAXY & ... & LINER \\
GTC\_17\_2 & 1237655495979106560 & 0.090 & 19.96 & 17.03 & -20.96 & GALAXY & STARFORMING & Composite \\\hline
GTC\_18\_1 & 1237655692474515647 & 0.175 & 20.23 & 16.58 & -23.35 & GALAXY & ... & LINER \\
GTC\_18\_2 & 1237655692474515648 & 0.174 & 20.56 & 17.86 & -21.80 & GALAXY & ... & Composite \\\hline
GTC\_19\_1 & 1237656239009104192 & 0.134 & 20.67 & 17.51 & -21.83 & GALAXY & ... & LINER \\
GTC\_19\_2 & 1237656239009104193 & 0.133 & 20.99 & 17.29 & -22.26 & GALAXY & ... & LINER \\\hline
GTC\_20\_1 & 1237656567582294307 & 0.027 & 18.58 & 15.54 & -21.74 & GALAXY & ... & Composite \\
GTC\_20\_2 & 1237656567582294308 & 0.026 & 18.77 & 16.24 & -20.42 & GALAXY & ... & Composite \\\hline
GTC\_21\_1 & 1237667253455356213 & 0.090 & 18.59 & 15.45 & -22.32 & GALAXY & ... & Composite \\
GTC\_21\_2 & \bf{1237667253455356214} & 0.091 & 22.51 & 19.36 & ... & GALAXY & ... & LINER \\\hline
GTC\_22\_1 & 1237656567585833094 & 0.057 & 17.99 & 14.97 & -22.26 & GALAXY & ... & LINER \\
GTC\_22\_2 & 1237656567585833097 & 0.057 & 19.58 & 15.8 & -21.95 & GALAXY & ... & Seyfert \\\hline
GTC\_23\_1 & 1237656906354393114 & 0.023 & 16.66 & 14.01 & -21.39 & GALAXY & ... & LINER \\
GTC\_23\_2 & 1237656906354393115 & 0.024 & 22.04 & 20.73 & -19.56 & GALAXY & STARFORMING & Composite \\\hline
GTC\_24\_1 & 1237657189817516083 & 0.139 & 20.71 & 17.63 & -22.55 & GALAXY & ... & LINER \\
GTC\_24\_2 & 1237657189817516085 & 0.138 & 22.1 & 18.3 & -21.23 & GALAXY & ... & Composite \\\hline
GTC\_25\_1 & 1237657221485625353 & 0.046 & 16.94 & 14.24 & -21.66 & GALAXY & BROADLINE & LINER \\
GTC\_25\_2 & 1237657221485559856 & 0.046 & 19.73 & 16.67 & -21.01 & GALAXY & ... & LINER \\\hline
GTC\_26\_1 & 1237657609104982532 & 0.153 & 20.17 & 17.06 & -22.36 & GALAXY & ... & Seyfert \\
GTC\_26\_2 & 1237657609104982533 & 0.153 & 21.66 & 17.89 & -21.68 & GALAXY & ... & LINER \\\hline
GTC\_27\_1 & 1237657612874940605 & 0.070 & 19.36 & 16.32 & -21.75 & GALAXY & ... & Composite \\
GTC\_27\_2 & 1237657612874940607 & 0.069 & 20.2 & 17.55 & -20.80 & GALAXY & ... & Composite \\\hline
GTC\_28\_1 & 1237657628432531602 & 0.078 & 18.91 & 15.74 & -22.40 & GALAXY & BROADLINE & LINER \\
GTC\_28\_2 & 1237657628432531603 & 0.078 & 20.36 & 17.13 & -21.32 & GALAXY & ... & Composite \\\hline
\\
GTC\_29\_1 & 1237657857148911724 & 0.060 & 19.35 & 16.73 & -20.17 & GALAXY & ... & LINER \\
GTC\_29\_2 & 1237657857148911725 & 0.060 & 18.6 & 16 & -21.21 & GALAXY & ... & Composite \\\hline
GTC\_30\_1 & 1237658300056338796 & 0.139 & 19.06 & 16.34 & -21.83 & GALAXY & STARFORMING & LINER \\
GTC\_30\_2 & 1237658300056338797 & 0.138 & 19.45 & 16.99 & -22.02 & GALAXY & ... & Composite \\\hline
GTC\_31\_1 & 1237658424090493080 & 0.072 & 19.34 & 16.62 & -21.51 & GALAXY & STARFORMING & Composite \\
GTC\_31\_2 & 1237658424090493081 & 0.071 & 19.15 & 16.23 & -21.00 & GALAXY & ... & Composite \\\hline
GTC\_32\_1 & 1237658611440222364 & 0.144 & 20.31 & 16.99 & -22.66 & GALAXY & ... & LINER \\
GTC\_32\_2 & 1237658611440222365 & 0.145 & 20.84 & 17.62 & -21.49 & GALAXY & ... & Composite \\\hline
GTC\_33\_1 & 1237658918533005504 & 0.113 & 20.16 & 17.27 & -21.40 & GALAXY & ... & Seyfert \\
GTC\_33\_2 & 1237658918533005505 & 0.113 & 20.56 & 17.87 & -20.91 & GALAXY & ... & Composite \\\hline
GTC\_34\_1 & 1237659119862939753 & 0.144 & 19.04 & 16.44 & -21.67 & GALAXY & STARFORMING & Composite \\
GTC\_34\_2 & 1237659119862939754 & 0.145 & 21.78 & 17.18 & -22.39 & GALAXY & ... & Seyfert \\\hline
GTC\_35\_1 & 1237659119864774675 & 0.099 & 19.37 & 15.84 & -22.40 & GALAXY & ... & Composite \\
GTC\_35\_2 & 1237659119864774676 & 0.100 & 20.51 & 17.43 & -21.18 & GALAXY & ... & Seyfert \\\hline
GTC\_36\_1 & 1237659132206448648 & 0.110 & 21.07 & 17.87 & -21.06 & GALAXY & ... & Seyfert \\
GTC\_36\_2 & 1237659132206448651 & 0.110 & 21.97 & 18.81 & -20.22 & GALAXY & ... & Seyfert \\\hline
GTC\_37\_1 & 1237660240313385134 & 0.142 & 21.95 & 18.23 & -21.13 & GALAXY & ... & LINER \\
GTC\_37\_2 & 1237660240313385135 & 0.142 & 20.55 & 17.75 & -20.68 & GALAXY & STARFORMING & Composite \\\hline
GTC\_38\_1 & 1237660240313712760 & 0.038 & 16.95 & 14.22 & -21.57 & GALAXY & STARFORMING & LINER \\
GTC\_38\_2 & 1237660240313712761 & 0.038 & 21.43 & 18.63 & -19.41 & GALAXY & ... & Composite \\\hline
GTC\_39\_1 & 1237660241385029806 & 0.074 & 19.27 & 16.32 & -21.34 & GALAXY & BROADLINE & LINER \\
GTC\_39\_2 & 1237660241385029810 & 0.073 & 20.12 & 17.54 & -21.77 & GALAXY & ... & Composite \\\hline
GTC\_40\_1 & 1237660669282812044 & 0.115 & 20.77 & 17.73 & -21.19 & GALAXY & ... & LINER \\
GTC\_40\_2 & 1237660669282812045 & 0.116 & 20.93 & 17.85 & -21.03 & GALAXY & ... & Seyfert \\\hline
GTC\_41\_1 & 1237660963472932915 & 0.048 & 18.15 & 15.01 & -22.37 & GALAXY & ... & LINER \\
GTC\_41\_2 & 1237660963472932916 & 0.048 & 18 & 15.57 & -21.94 & GALAXY & STARFORMING & Composite \\\hline
GTC\_42\_1 & 1237661086415585604 & 0.047 & 15.96 & 13.14 & -22.60 & GALAXY & ... & LINER \\
GTC\_42\_2 & 1237661086415585607 & 0.048 & 19.52 & 15.61 & -22.33 & GALAXY & ... & Composite \\\hline
GTC\_43\_1 & 1237661137963254106 & 0.129 & 19.75 & 16.6 & -22.43 & GALAXY & ... & LINER \\
GTC\_43\_2 & 1237661137963254107 & 0.129 & 20.98 & 17.83 & -21.78 & GALAXY & ... & LINER \\\hline
GTC\_44\_1 & 1237661151377424586 & 0.104 & 20.6 & 16.58 & -21.65 & GALAXY & ... & Composite \\
GTC\_44\_2 & 1237661151377424587 & 0.104 & 20.34 & 17.73 & -20.64 & GALAXY & ... & Seyfert \\\hline
GTC\_45\_1 & 1237661352706179107 & 0.082 & 18.21 & 15.82 & -21.73 & GALAXY & ... & Composite \\
GTC\_45\_2 & 1237661352706179108 & 0.082 & 19.85 & 16.8 & -21.47 & GALAXY & ... & Composite \\\hline
GTC\_46\_1 & 1237661382778093715 & 0.069 & 18.16 & 15.3 & -21.95 & GALAXY & ... & Composite \\
GTC\_46\_2 & 1237661382778093716 & 0.068 & 18.86 & 16.04 & -21.48 & GALAXY & BROADLINE & Composite \\\hline
GTC\_47\_1 & 1237661418748444829 & 0.107 & 19.09 & 16.09 & -22.00 & GALAXY & ... & LINER \\
GTC\_47\_2 & 1237661418748444830 & 0.107 & 19.94 & 16.95 & -21.22 & GALAXY & ... & Seyfert \\\hline
GTC\_48\_1 & 1237661812274233471 & 0.064 & 17.55 & 15.59 & -22.10 & GALAXY & STARBURST & Composite \\
GTC\_48\_2 & 1237661812274233472 & 0.064 & 19.4 & 16.41 & -21.38 & GALAXY & ... & LINER \\\hline
GTC\_49\_1 & 1237661813348106331 & 0.123 & 19.74 & 16.67 & -22.04 & GALAXY & ... & Seyfert \\
GTC\_49\_2 & 1237661813348106332 & 0.124 & 20.7 & 17.74 & -21.03 & GALAXY & ... & Composite \\\hline
GTC\_50\_1 & 1237661850411204729 & 0.076 & 18.19 & 15.75 & -21.96 & GALAXY & STARFORMING & Composite \\
GTC\_50\_2 & 1237661850411204730 & 0.075 & 19.42 & 16.5 & -21.49 & GALAXY & ... & Seyfert \\\hline
GTC\_51\_1 & 1237661871330361452 & 0.112 & 19.47 & 16.36 & -22.32 & GALAXY & ... & LINER \\
GTC\_51\_2 & 1237661871330361453 & 0.112 & 18.84 & 15.77 & -22.72 & GALAXY & ... & Composite \\\hline
GTC\_52\_1 & 1237661957762842806 & 0.133 & 20.94 & 17.21 & -22.04 & GALAXY & ... & Composite \\
GTC\_52\_2 & 1237661957762842807 & 0.133 & 20.79 & 17.34 & -21.58 & GALAXY & ... & Seyfert \\\hline
GTC\_53\_1 & 1237661966886633584 & 0.131 & 20.61 & 17.27 & -21.89 & GALAXY & ... & LINER \\
GTC\_53\_2 & 1237661966886633585 & 0.131 & 20.69 & 17.6 & -21.82 & GALAXY & ... & LINER \\\hline
GTC\_54\_1 & 1237661971721945220 & 0.073 & 19.22 & 16.23 & -21.74 & GALAXY & ... & Seyfert \\
GTC\_54\_2 & 1237661971721945221 & 0.074 & 18.43 & 15.72 & -21.23 & GALAXY & ... & LINER \\\hline
GTC\_55\_1 & 1237662193453760669 & 0.171 & 20.6 & 17.6 & -22.20 & GALAXY & ... & Composite \\
GTC\_55\_2 & 1237662193453760670 & 0.173 & 20.96 & 17.62 & -22.01 & GALAXY & ... & Seyfert \\\hline
GTC\_56\_1 & 1237662194541002843 & 0.073 & 18.36 & 15.47 & -22.48 & GALAXY & ... & LINER \\
GTC\_56\_2 & 1237662194541002844 & 0.074 & 18.4 & 15.8 & -21.54 & GALAXY & ... & Seyfert \\\hline
\\
GTC\_57\_1 & 1237662198280552769 & 0.172 & 21.03 & 17.62 & -22.17 & GALAXY & ... & LINER \\
GTC\_57\_2 & 1237662198280552768 & 0.172 & 21.25 & 17.69 & -22.25 & GALAXY & ... & LINER \\\hline
GTC\_58\_1 & 1237662225666932750 & 0.099 & 20 & 16.83 & -21.67 & GALAXY & ... & Composite \\
GTC\_58\_2 & 1237662225666932751 & 0.099 & 20.92 & 18.21 & -20.28 & GALAXY & ... & Composite \\\hline
GTC\_59\_1 & 1237662335178178801 & 0.065 & 18.45 & 15.57 & -22.17 & GALAXY & ... & Seyfert \\
GTC\_59\_2 & 1237662335178178802 & 0.065 & 19.85 & 17.05 & -20.85 & GALAXY & ... & Seyfert \\\hline
GTC\_60\_1 & 1237662336792920258 & 0.069 & 18.99 & 16.25 & -21.25 & GALAXY & ... & LINER \\
GTC\_60\_2 & 1237662336792920257 & 0.070 & 18.33 & 15.67 & -21.27 & GALAXY & STARFORMING & Composite \\\hline
GTC\_61\_1 & 1237662474231021741 & 0.117 & 19.12 & 16.09 & -22.85 & GALAXY & ... & LINER \\
GTC\_61\_2 & 1237662474231021742 & 0.117 & 20.26 & 17.23 & -20.96 & GALAXY & ... & Composite \\\hline
GTC\_62\_1 & 1237662498390737019 & 0.154 & 19.87 & 16.69 & -22.60 & GALAXY & ... & LINER \\
GTC\_62\_2 & 1237662498390737020 & 0.153 & 20.01 & 16.95 & -22.72 & GALAXY & ... & LINER \\\hline
GTC\_63\_1 & 1237662200436752504 & 0.105 & 19.28 & 16.09 & -22.49 & GALAXY & ... & LINER \\
GTC\_63\_2 & 1237662636375867758 & 0.105 & 19.86 & 16.8 & -21.93 & GALAXY & STARFORMING & Composite \\\hline
GTC\_64\_1 & 1237662640123609565 & 0.059 & 18.42 & 15.48 & -21.90 & GALAXY & ... & Composite \\
GTC\_64\_2 & 1237662640123609566 & 0.060 & 24.26 & 21.75 & -20.71 & GALAXY & ... & Composite \\\hline
GTC\_65\_1 & 1237663463145537628 & 0.084 & 17.15 & 13.92 & -23.95 & GALAXY & ... & Seyfert \\
GTC\_65\_2 & 1237663463145537634 & 0.084 & 28.58 & 16.97 & -24.36 & GALAXY & ... & LINER \\\hline
GTC\_66\_1 & 1237663543683318057 & 0.084 & 19.87 & 17.1 & -20.45 & GALAXY & ... & Composite \\
GTC\_66\_2 & 1237663543683318058 & 0.085 & 20.4 & 17.93 & -20.57 & GALAXY & STARFORMING & Composite \\\hline
GTC\_67\_1 & 1237663782589432032 & 0.139 & 20.97 & 17.8 & -21.66 & GALAXY & ... & Composite \\
GTC\_67\_2 & 1237663782589432033 & 0.139 & 20.78 & 17.8 & -21.44 & GALAXY & ... & Composite \\\hline
GTC\_68\_1 & 1237663783122370706 & 0.186 & 20.76 & 17.25 & -22.73 & GALAXY & ... & LINER \\
GTC\_68\_2 & 1237663783122370707 & 0.183 & 22.18 & 18.77 & -21.18 & GALAXY & ... & LINER \\\hline
GTC\_69\_1 & 1237663783131283514 & 0.089 & 17.89 & 16.52 & -20.21 & GALAXY & STARBURST & Composite \\
GTC\_69\_2 & 1237663783131283515 & 0.089 & 19.53 & 16.59 & -21.31 & GALAXY & ... & Composite \\\hline
GTC\_70\_1 & 1237663783136985126 & 0.120 & 20.64 & 17.85 & -20.38 & GALAXY & ... & Composite \\
GTC\_70\_2 & 1237663783136985125 & 0.121 & 20.12 & 17.05 & -21.85 & GALAXY & ... & Seyfert \\\hline
GTC\_71\_1 & 1237663783674183846 & 0.112 & 20.19 & 17.25 & -21.09 & GALAXY & STARFORMING & Composite \\
GTC\_71\_2 & 1237663783674183847 & 0.111 & 19.92 & 18.01 & -20.16 & GALAXY & STARFORMING & Composite \\\hline
GTC\_72\_1 & 1237663784195653787 & 0.079 & 20.77 & 17.53 & -21.71 & GALAXY & ... & LINER \\
GTC\_72\_2 & 1237663784195653788 & 0.079 & 19.68 & 17.67 & -18.91 & GALAXY & ... & Composite \\\hline
GTC\_73\_1 & 1237663784200831188 & 0.073 & 19.51 & 16.65 & -21.10 & GALAXY & ... & LINER \\
GTC\_73\_2 & 1237663784200831189 & 0.074 & 19.8 & 17.48 & -20.61 & GALAXY & STARFORMING & Composite \\\hline
GTC\_74\_1 & 1237663784739995886 & 0.115 & 19 & 16.79 & -20.96 & GALAXY & STARFORMING & Composite \\
GTC\_74\_2 & 1237663784739995887 & 0.116 & 20.39 & 17.38 & -21.84 & GALAXY & ... & Composite \\\hline
GTC\_75\_1 & 1237664291548823599 & 0.041 & 17.88 & 15.17 & -21.74 & GALAXY & BROADLINE & LINER \\
GTC\_75\_2 & 1237664291548823600 & 0.041 & 16.38 & 13.92 & -21.74 & GALAXY & ... & LINER \\\hline
GTC\_76\_1 & 1237664667352039631 & 0.084 & 20.13 & 17.08 & -21.09 & GALAXY & ... & LINER \\
GTC\_76\_2 & 1237664667352039632 & 0.085 & 19.1 & 16.84 & -21.23 & GALAXY & STARFORMING & LINER \\\hline
GTC\_77\_1 & 1237664817676484746 & 0.215 & 19.86 & 17.79 & -22.19 & GALAXY & STARFORMING & Composite \\
GTC\_77\_2 & 1237664817676484748 & 0.216 & 20.71 & 17.84 & -21.76 & GALAXY & STARFORMING & Composite \\\hline
GTC\_78\_1 & 1237664871900577858 & 0.049 & 19.33 & 16.45 & -21.01 & GALAXY & ... & LINER \\
GTC\_78\_2 & 1237664871900577859 & 0.051 & 19.73 & 17.07 & -21.50 & GALAXY & ... & Seyfert \\\hline
GTC\_79\_1 & 1237664877803077714 & 0.116 & 20.7 & 17.59 & -21.25 & GALAXY & ... & LINER \\
GTC\_79\_2 & 1237664877803077715 & 0.118 & 20.59 & 17.43 & -21.80 & GALAXY & ... & LINER \\\hline
GTC\_80\_1 & 1237664877807927453 & 0.206 & 21.04 & 17.68 & -22.63 & GALAXY & ... & Seyfert \\
GTC\_80\_2 & 1237664877807927454 & 0.207 & 22.19 & 17.62 & -22.57 & GALAXY & ... & LINER \\\hline
GTC\_81\_1 & 1237665103283814519 & 0.075 & 19.44 & 16.54 & -21.57 & GALAXY & ... & LINER \\
GTC\_81\_2 & 1237665103283814520 & 0.075 & 18.36 & 15.57 & -21.89 & GALAXY & ... & LINER \\\hline
GTC\_82\_1 & 1237665129610936526 & 0.154 & 21.23 & 17.56 & -21.86 & GALAXY & ... & LINER \\
GTC\_82\_2 & 1237665129610936527 & 0.153 & 21.66 & 17.89 & -21.23 & GALAXY & STARFORMING & Composite \\\hline
GTC\_83\_1 & 1237665328779100316 & 0.154 & 21.21 & 17.63 & -21.96 & GALAXY & ... & LINER \\
GTC\_83\_2 & 1237665328779100317 & 0.155 & 20.06 & 16.98 & -22.56 & GALAXY & ... & Seyfert \\\hline
GTC\_84\_1 & 1237665330927435873 & 0.088 & 19.2 & 16.48 & -21.65 & GALAXY & STARBURST & Composite \\
GTC\_84\_2 & 1237665330927435875 & 0.088 & 20.68 & 18.25 & -19.99 & GALAXY & ... & Seyfert \\\hline
\\
GTC\_85\_1 & 1237665331455459378 & 0.072 & 18.09 & 15.84 & -22.28 & GALAXY & STARFORMING & Seyfert \\
GTC\_85\_2 & 1237665331455459379 & 0.073 & 18.13 & 17.16 & -20.11 & GALAXY & STARFORMING & Composite \\\hline
GTC\_86\_1 & 1237665331457425465 & 0.128 & 19.99 & 16.62 & -22.59 & GALAXY & ... & LINER \\
GTC\_86\_2 & 1237665331457425466 & 0.127 & 20.09 & 16.8 & -22.65 & GALAXY & ... & LINER \\\hline
GTC\_87\_1 & 1237665428091240506 & 0.140 & 18.97 & 16.27 & -22.55 & GALAXY & ... & Seyfert \\
GTC\_87\_2 & 1237665428091240507 & 0.141 & 21.26 & 18.32 & -20.88 & GALAXY & ... & Composite \\\hline
GTC\_88\_1 & 1237665429713780966 & 0.088 & 20.43 & 17.5 & -20.62 & GALAXY & ... & Composite \\
GTC\_88\_2 & 1237665429713846350 & 0.090 & 19.39 & 16.57 & -21.76 & GALAXY & ... & LINER \\\hline
GTC\_89\_1 & 1237665440439336990 & 0.065 & 17.45 & 14.57 & -22.50 & GALAXY & ... & Composite \\
GTC\_89\_2 & 1237665440439336991 & 0.061 & 17.74 & 14.82 & -22.30 & GALAXY & ... & LINER \\\hline
GTC\_90\_1 & 1237665531723579489 & 0.132 & 19.47 & 16.37 & -22.38 & GALAXY & ... & LINER \\
GTC\_90\_2 & 1237665531723579490 & 0.130 & 20.12 & 17.16 & -21.66 & GALAXY & ... & LINER \\\hline
GTC\_91\_1 & 1237665566082269200 & 0.094 & 18.86 & 15.76 & -22.41 & GALAXY & ... & LINER \\
GTC\_91\_2 & 1237665566082269201 & 0.094 & 20.25 & 17.2 & -21.29 & GALAXY & ... & Composite \\\hline
GTC\_92\_1 & 1237666185634447638 & 0.121 & 21.03 & 17.64 & -21.49 & GALAXY & ... & LINER \\
GTC\_92\_2 & 1237666185634447639 & 0.121 & 20.97 & 17.3 & -22.01 & GALAXY & ... & LINER \\\hline
GTC\_93\_1 & 1237666301627924604 & 0.156 & 23.1 & 19.58 & -20.87 & GALAXY & ... & Composite \\
GTC\_93\_2 & 1237666301627924605 & 0.156 & 19.48 & 16.97 & -21.76 & GALAXY & STARFORMING & LINER \\\hline
GTC\_94\_1 & 1237666407381598382 & 0.080 & 19.72 & 17.08 & -21.20 & GALAXY & STARFORMING & Composite \\
GTC\_94\_2 & 1237666407381598383 & 0.080 & 18.3 & 16.49 & -19.44 & GALAXY & ... & Composite \\\hline
GTC\_95\_1 & 1237667109574279383 & 0.201 & 21.18 & 17.51 & -22.75 & GALAXY & ... & LINER \\
GTC\_95\_2 & 1237667109574279384 & 0.200 & 21.47 & 17.79 & -22.40 & GALAXY & ... & Composite \\\hline
GTC\_96\_1 & 1237667209974710300 & 0.078 & 19.19 & 17.4 & -20.40 & GALAXY & STARBURST & Composite \\
GTC\_96\_2 & 1237667209974710301 & 0.078 & 19.49 & 16.83 & -21.00 & GALAXY & STARFORMING & Composite \\\hline
GTC\_97\_1 & 1237667212115050866 & 0.172 & 20.88 & 17.51 & -22.48 & GALAXY & ... & LINER \\
GTC\_97\_2 & 1237667212115050867 & 0.173 & 21.84 & 18.38 & -21.42 & GALAXY & ... & Composite \\\hline
GTC\_98\_1 & 1237667212136284360 & 0.138 & 21.01 & 17.71 & -21.44 & GALAXY & ... & LINER \\
GTC\_98\_2 & 1237667212136284359 & 0.130 & 19.49 & 17.03 & -21.30 & GALAXY & ... & Seyfert \\\hline
GTC\_99\_1 & 1237667253453324693 & 0.141 & 19.56 & 16.64 & -21.99 & GALAXY & ... & LINER \\
GTC\_99\_2 & 1237667253453324694 & 0.141 & 21.16 & 17.11 & -22.04 & GALAXY & ... & LINER \\\hline
GTC\_100\_1 & 1237667254546006103 & 0.145 & 19.25 & 17.88 & -20.04 & GALAXY & STARFORMING & Composite \\
GTC\_100\_2 & 1237667254546006104 & 0.145 & 20.88 & 17.84 & -20.48 & GALAXY & ... & LINER \\\hline
GTC\_101\_1 & 1237667323250213040 & 0.038 & 17.99 & 15.28 & -21.45 & GALAXY & STARFORMING & Composite \\
GTC\_101\_2 & 1237667323250213041 & 0.038 & 19.77 & 18 & -18.23 & GALAXY & ... & Composite \\\hline
GTC\_102\_1 & 1237667323796127885 & 0.083 & 18.76 & 16.18 & -21.58 & GALAXY & ... & LINER \\
GTC\_102\_2 & 1237667323796127886 & 0.083 & 20.04 & 17.47 & -20.71 & GALAXY & STARFORMING & Composite \\\hline
GTC\_103\_1 & 1237667442437914759 & 0.079 & 19.57 & 16.64 & -21.55 & GALAXY & ... & LINER \\
GTC\_103\_2 & 1237667442437914760 & 0.080 & 19.25 & 16.73 & -21.25 & GALAXY & STARFORMING & Composite \\\hline
GTC\_104\_1 & 1237667549268148435 & 0.118 & 19.11 & 16.73 & -21.37 & GALAXY & STARFORMING & Composite \\
GTC\_104\_2 & 1237667549268213959 & 0.116 & 18.99 & 16.43 & -21.31 & GALAXY & STARFORMING & Composite \\\hline
GTC\_105\_1 & 1237667733989949671 & 0.090 & 20.34 & 17.36 & -21.04 & GALAXY & ... & Composite \\
GTC\_105\_2 & 1237667733989949676 & 0.089 & 20.39 & 17.31 & -21.04 & GALAXY & ... & Seyfert \\\hline
GTC\_106\_1 & 1237667734526886197 & 0.119 & 20.61 & 17.65 & -21.30 & GALAXY & ... & Composite \\
GTC\_106\_2 & 1237667734526886198 & 0.118 & 20.44 & 17.08 & -21.71 & GALAXY & ... & Seyfert \\\hline
GTC\_107\_1 & 1237667736133828618 & 0.096 & 18.9 & 15.97 & -22.06 & GALAXY & ... & Seyfert \\
GTC\_107\_2 & 1237667736133828619 & 0.097 & 18.99 & 16.07 & -22.27 & GALAXY & ... & LINER \\\hline
GTC\_108\_1 & 1237667781774934219 & 0.102 & 18.81 & 15.87 & -21.80 & GALAXY & ... & Composite \\
GTC\_108\_2 & 1237667781774934220 & 0.103 & 20.42 & 17.3 & -21.21 & GALAXY & ... & Seyfert \\\hline
GTC\_109\_1 & 1237667783883817113 & 0.148 & 24.68 & 17.24 & -22.41 & GALAXY & ... & LINER \\
GTC\_109\_2 & 1237667783883817114 & 0.149 & 20.12 & 17.55 & -22.02 & GALAXY & ... & Seyfert \\\hline
GTC\_110\_1 & 1237667783929299179 & 0.051 & 17.49 & 14.64 & -22.07 & GALAXY & BROADLINE & LINER \\
GTC\_110\_2 & 1237667783929299180 & 0.050 & 18.31 & 15.66 & -21.49 & GALAXY & BROADLINE & LINER \\\hline
GTC\_111\_1 & 1237667910069190795 & 0.077 & 18.24 & 16.11 & -21.18 & GALAXY & STARFORMING & Composite \\
GTC\_111\_2 & 1237667910069190796 & 0.077 & 18.33 & 16.16 & -22.39 & GALAXY & STARFORMING & Composite \\\hline
GTC\_112\_1 & 1237667916494012508 & 0.066 & 18.51 & 15.76 & -21.44 & GALAXY & ... & Seyfert \\
GTC\_112\_2 & 1237667916494012509 & 0.065 & 18.69 & 15.76 & -21.63 & GALAXY & ... & LINER \\\hline
\\
GTC\_113\_1 & 1237668271362211975 & 0.088 & 19.37 & 16.28 & -22.22 & GALAXY & ... & LINER \\
GTC\_113\_2 & 1237668271362211976 & 0.089 & 19.78 & 17.1 & -20.48 & GALAXY & ... & LINER \\\hline
GTC\_114\_1 & 1237668298204446938 & 0.119 & 18.81 & 17.09 & -21.83 & GALAXY & STARBURST & Composite \\
GTC\_114\_2 & 1237668298204446939 & 0.119 & 20.7 & 16.93 & -22.08 & GALAXY & STARFORMING & Composite \\\hline
GTC\_115\_1 & 1237668589722534078 & 0.153 & 20.55 & 17.59 & -21.62 & GALAXY & STARFORMING & Composite \\
GTC\_115\_2 & 1237668589722534074 & 0.152 & 21.45 & 17.98 & -21.72 & GALAXY & ... & Composite \\\hline
GTC\_116\_1 & 1237668623551103255 & 0.216 & 20.18 & 17.72 & -22.03 & GALAXY & ... & Composite \\
GTC\_116\_2 & 1237668623551103256 & 0.216 & 23.16 & 18.94 & -23.00 & GALAXY & ... & Composite \\\hline
GTC\_117\_1 & 1237668672938115205 & 0.136 & 19.77 & 16.4 & -22.48 & GALAXY & ... & LINER \\
GTC\_117\_2 & 1237668672938115206 & 0.135 & 21.01 & 17.89 & -21.07 & GALAXY & ... & Composite \\\hline
GTC\_118\_1 & 1237670964313456852 & 0.076 & 18.02 & 15.16 & -22.48 & GALAXY & ... & LINER \\
GTC\_118\_2 & 1237670964313456853 & 0.078 & 18.52 & 15.61 & -22.12 & GALAXY & ... & LINER \\\hline
GTC\_119\_1 & 1237670965383266430 & 0.031 & 17.81 & 15.04 & -21.53 & GALAXY & ... & LINER \\
GTC\_119\_2 & 1237670965383266433 & 0.030 & 18.18 & 15.57 & -21.36 & GALAXY & ... & Composite \\\hline
GTC\_120\_1 & 1237671931208925416 & 0.073 & 18.22 & 15.26 & -22.37 & GALAXY & ... & LINER \\
GTC\_120\_2 & 1237671931208925417 & 0.073 & 19.75 & 16.78 & -21.46 & GALAXY & ... & LINER \\\hline
GTC\_121\_1 & 1237673807580693064 & 0.144 & 20.13 & 16.39 & -22.88 & GALAXY & ... & LINER \\
GTC\_121\_2 & 1237673807580693069 & 0.140 & 20.46 & 17.37 & -21.18 & GALAXY & ... & Composite \\\hline
GTC\_122\_1 & 1237673808116122327 & 0.100 & 18.35 & 16.19 & -21.85 & GALAXY & STARFORMING & Composite \\
GTC\_122\_2 & 1237673808116122329 & 0.099 & 19.45 & 16.4 & -21.40 & GALAXY & ... & LINER \\\hline
GTC\_123\_1 & 1237674649921061067 & 0.096 & 19.22 & 16.11 & -22.29 & GALAXY & ... & Composite \\
GTC\_123\_2 & 1237674649921061068 & 0.093 & 19.21 & 16.32 & -21.94 & GALAXY & ... & Composite \\\hline
GTC\_124\_1 & 1237657587096289353 & 0.025 & 16.62 & 13.89 & -21.51 & GALAXY & ... & Seyfert \\
GTC\_124\_2 & \bf{1237657587096289354} & 0.025 & 20.44 & 18.21 & ... & GALAXY & ... & Seyfert \\\hline
GTC\_125\_1 & 1237666210341978625 & 0.060 & 19.28 & 16.97 & -20.89 & GALAXY & STARFORMING & Composite \\
GTC\_125\_2 & 1237666210341978627 & 0.061 & 18.89 & 16.36 & -21.33 & GALAXY & STARFORMING & Composite \\\hline
GTC\_126\_1 & 1237662335180341316 & 0.064 & 18.23 & 15.84 & -21.52 & GALAXY & STARFORMING & Composite \\
GTC\_126\_2 & \bf{1237662335180341327} & 0.064 & 22.59 & 21.88 & ... & GALAXY & STARFORMING & Composite \\\hline
GTC\_127\_1 & 1237659327092752463 & 0.149 & 18.81 & 17.52 & -22.07 & QSO & STARBURST & Seyfert \\
GTC\_127\_2 & 1237659327092752464 & 0.150 & 20.58 & 17.62 & -21.56 & GALAXY & STARFORMING & Composite \\\hline
GTC\_128\_1 & 1237667915957665832 & 0.065 & 17.86 & 14.74 & -22.90 & GALAXY & ... & Composite \\
GTC\_128\_2 & 1237667915957665835 & 0.065 & 18.1 & 15.24 & -21.86 & GALAXY & AGN & Composite \\\hline
GTC\_129\_1 & 1237668588651282620 & 0.074 & 18.39 & 15.57 & -21.94 & GALAXY & ... & LINER \\
GTC\_129\_2 & 1237668588651282621 & 0.074 & 18.24 & 15.76 & -21.15 & GALAXY & AGN & LINER \\\hline
GTC\_130\_1 & 1237671262267506768 & 0.052 & 16.96 & 14.02 & -22.83 & GALAXY & AGN & LINER \\
GTC\_130\_2 & 1237671262267506767 & 0.053 & 19.53 & 16.73 & -20.68 & GALAXY & ... & LINER \\\hline
GTC\_131\_1 & 1237666227490586650 & 0.103 & 18.23 & 16.29 & -22.30 & GALAXY & STARBURST & Composite \\
GTC\_131\_2 & 1237666227490586649 & 0.104 & 20.69 & 17.95 & -21.58 & QSO & STARBURST & Composite \\\hline
GTC\_132\_1 & 1237666228063371544 & 0.100 & 20.39 & 16.97 & -21.91 & GALAXY & AGN & Seyfert \\
GTC\_132\_2 & 1237666228063371545 & 0.099 & 18.16 & 16.08 & -22.06 & GALAXY & STARFORMING & Composite \\\hline
GTC\_133\_1 & 1237667252917895553 & 0.153 & 20.31 & 16.48 & -23.22 & GALAXY & ... & LINER \\
GTC\_133\_2 & 1237667252917895554 & 0.156 & 18.53 & 16.94 & -22.36 & QSO & STARBURST & Composite \\\hline
GTC\_134\_1 & 1237664667360755820 & 0.152 & 19.89 & 18.03 & -22.35 & QSO & STARBURST & Composite \\
GTC\_134\_2 & 1237664667360755822 & 0.153 & 20.87 & 19.29 & -18.30 & GALAXY & AGN & Composite \\\hline
GTC\_135\_1 & 1237660240314433647 & 0.156 & 20.74 & 17.5 & -22.79 & GALAXY & ... & Seyfert \\
GTC\_135\_2 & 1237660240314433648 & 0.158 & 21.46 & 18.07 & -21.43 & GALAXY & AGN & Seyfert \\\hline
GTC\_136\_1 & 1237655502962688659 & 0.030 & 17.75 & 15.16 & -20.91 & GALAXY & AGN & LINER \\
GTC\_136\_2 & 1237655502962688660 & 0.029 & 20.5 & 18.1 & -22.72 & GALAXY & STARFORMING & Composite \\\hline
GTC\_137\_1 & 1237655465916104829 & 0.132 & 20.15 & 17.28 & -21.32 & GALAXY & AGN & LINER \\
GTC\_137\_2 & 1237655465916104830 & 0.132 & 21.25 & 17.5 & -22.04 & GALAXY & ... & LINER \\\hline
GTC\_138\_1 & 1237653616938516591 & 0.099 & 19.02 & 16.19 & -21.58 & GALAXY & AGN & LINER \\
GTC\_138\_2 & 1237653616938516592 & 0.099 & 19.86 & 16.84 & -21.23 & GALAXY & ... & LINER \\\hline
GTC\_139\_1 & 1237652600635851575 & 0.175 & 19.61 & 17.01 & -22.90 & GALAXY & AGN & LINER \\
GTC\_139\_2 & 1237652600635851576 & 0.178 & 20.45 & 17.13 & -22.72 & GALAXY & ... & LINER \\\hline
GTC\_140\_1 & 1237651250968461337 & 0.223 & 18.97 & 18.39 & -21.39 & QSO & STARBURST & Composite \\
GTC\_140\_2 & 1237651250968461338 & 0.223 & 24.15 & 20.53 & -20.80 & GALAXY & STARBURST & Composite \\\hline
\\
GTC\_141\_1 & 1237648721245503520 & 0.047 & 18.83 & 17.16 & -21.21 & GALAXY & STARBURST & Composite \\
GTC\_141\_2 & 1237648721245503521 & 0.048 & 17.92 & 15.52 & -21.30 & GALAXY & AGN & Seyfert \\\hline
GTC\_142\_1 & 1237648722312888376 & 0.138 & 18.92 & 17.16 & -21.79 & GALAXY & STARBURST & Seyfert \\
GTC\_142\_2 & 1237648722312888377 & 0.138 & 19.79 & 17.08 & -20.79 & GALAXY & AGN & Seyfert \\\hline
GTC\_143\_1 & 1237654382514471111 & 0.060 & 18.64 & 16.18 & -20.81 & GALAXY & AGN & Seyfert \\
GTC\_143\_2 & 1237654382514471112 & 0.060 & 19.48 & 17.05 & -20.88 & GALAXY & STARFORMING & Composite \\\hline
GTC\_144\_1 & 1237657191979286606 & 0.087 & 19.23 & 16.06 & -21.87 & QSO & STARBURST & Seyfert \\
GTC\_144\_2 & 1237657191979286607 & 0.087 & 19.26 & 16.35 & -21.78 & GALAXY & ... & LINER \\\hline
GTC\_145\_1 & 1237659326018355482 & 0.105 & 19.69 & 16.7 & -21.64 & GALAXY & ... & Seyfert \\
GTC\_145\_2 & 1237659326018355483 & 0.107 & 20.48 & 17.54 & -21.64 & GALAXY & AGN & Seyfert \\\hline
GTC\_146\_1 & 1237661465992626516 & 0.063 & 19.53 & 16.54 & -21.64 & GALAXY & BROADLINE & LINER \\
GTC\_146\_2 & 1237661465992626517 & 0.065 & 19.36 & 16.22 & -21.65 & GALAXY & AGN & LINER \\\hline
GTC\_147\_1 & 1237661812275347597 & 0.091 & 19.26 & 16.87 & -21.56 & GALAXY & AGN & Seyfert \\
GTC\_147\_2 & 1237661812275347598 & 0.092 & 20.65 & 17.72 & -21.17 & GALAXY & ... & Composite \\\hline
GTC\_148\_1 & 1237661813885436021 & 0.065 & 19.04 & 16.11 & -21.43 & GALAXY & ... & LINER \\
GTC\_148\_2 & 1237661813885436022 & 0.065 & 19.33 & 16.79 & -21.20 & GALAXY & AGN & Seyfert \\\hline
GTC\_149\_1 & 1237662224058089529 & 0.087 & 19.62 & 16.6 & -21.79 & GALAXY & ... & Composite \\
GTC\_149\_2 & 1237662224058024149 & 0.088 & 19.78 & 16.77 & -21.21 & GALAXY & AGN & LINER \\\hline
GTC\_150\_1 & 1237663542609117199 & 0.128 & 20.62 & 17.38 & -22.12 & QSO & STARBURST & Seyfert \\
GTC\_150\_2 & 1237663542609117200 & 0.128 & 20.71 & 17.11 & -21.51 & GALAXY & ... & Composite \\\hline
GTC\_151\_1 & 1237657192517926989 & 0.105 & 20.54 & 17.34 & -21.28 & GALAXY & ... & LINER \\
GTC\_151\_2 & 1237657192517926990 & 0.106 & 22 & 18.94 & -20.24 & GALAXY & ... & LINER \\
GTC\_151\_3 & 1237657192517926993 & 0.106 & 19.77 & 16.88 & -21.83 & GALAXY & ... & Seyfert \\\hline
GTC\_152\_1 & 1237661064950710579 & 0.126 & 19.37 & 16.03 & -22.85 & GALAXY & ... & Composite \\
GTC\_152\_2 & 1237661064950710580 & 0.125 & 19.63 & 16.71 & -22.17 & GALAXY & ... & LINER \\
GTC\_152\_3 & \bf{1237661064950710581} & 0.127 & 21.28 & 18.91 & ... & GALAXY & ... & ... \\\hline
GTC\_153\_1 & 1237663275777392799 & 0.094 & 21.26 & 18.1 & -20.84 & GALAXY & ... & Composite \\
GTC\_153\_2 & 1237663275777392800 & 0.095 & 19.29 & 16.13 & -22.01 & GALAXY & ... & Seyfert \\
GTC\_153\_3 & 1237663275777392798 & 0.091 & 20.77 & 17.53 & -21.43 & GALAXY & ... & SF \\\hline
GTC\_154\_1 & 1237663542608920721 & 0.144 & 21.47 & 17.59 & -22.03 & GALAXY & ... & LINER \\
GTC\_154\_2 & 1237663542608920724 & 0.144 & 24.33 & 20.2 & -21.12 & GALAXY & ... & Seyfert \\
GTC\_154\_3 & 1237663542608920720 & 0.144 & 22.06 & 17.82 & -21.73 & GALAXY & ... & SF \\\hline
GTC\_155\_1 & 1237663783126237343 & 0.063 & 19.48 & 16.54 & -21.29 & GALAXY & ... & Composite \\
GTC\_155\_2 & 1237663783126237344 & 0.064 & 19.51 & 16.44 & -21.49 & GALAXY & ... & LINER \\
GTC\_155\_3 & 1237663783126237342 & 0.064 & 18.34 & 15.34 & -22.12 & GALAXY & ... & SF \\\hline
GTC\_156\_1 & 1237665564997714246 & 0.260 & 21.57 & 18.18 & -22.63 & GALAXY & ... & LINER \\
GTC\_156\_2 & 1237665564997779740 & 0.254 & 21.74 & 16.69 & -24.02 & GALAXY & ... & Composite \\
GTC\_156\_3 & 1237665564997714249 & 0.254 & 24.19 & 18.2 & -22.67 & GALAXY & ... & ... \\\hline
GTC\_157\_1 & 1237668299280089194 & 0.063 & 18.38 & 15.41 & -22.07 & GALAXY & ... & LINER \\
GTC\_157\_2 & 1237668299280089196 & 0.063 & 19.45 & 16.93 & -21.01 & GALAXY & STARFORMING & Composite \\
GTC\_157\_3 & 1237668299280089195 & 0.062 & 18.57 & 15.73 & -21.64 & GALAXY & ... & ... \\\hline
GTC\_158\_1 & 1237667736133107729 & 0.102 & 19.05 & 16.05 & -22.55 & GALAXY & ... & Seyfert \\
GTC\_158\_2 & 1237667736133107730 & 0.103 & 18.99 & 15.96 & -22.82 & GALAXY & ... & LINER \\
GTC\_158\_3 & 1237667736133107731 & 0.103 & 19.98 & 16.84 & -21.44 & GALAXY & BROADLINE & ... \\\hline
GTC\_159\_1 & 1237666301093675376 & 0.086 & 19.37 & 16.76 & -21.31 & QSO & BROADLINE & LINER \\
GTC\_159\_2 & 1237666301093675377 & 0.087 & 20.96 & 18.27 & -19.72 & GALAXY & ... & LINER \\
GTC\_159\_3 & 1237666301093675378 & 0.088 & 19.88 & 17.87 & -20.96 & GALAXY & STARBURST & SF \\
GTC\_159\_4 & 1237666301093675379 & 0.086 & 20.92 & 18.99 & -19.24 & GALAXY & STARFORMING & SF \\ \hline
\caption{\mybf{Data from SDSS DR16 for all the DAGN objects identified by GOTHIC. There are 4 objects (marked in bold) whose absolute r-band magnitude values are missing. In SDSS object explorer, one of them is flagged to have unreliable photometric data and the others have related photometric flags which could explain their missing data.}}
\end{longtable}

\twocolumn

\bibliographystyle{mnras}
\bibliography{bibliography}

\ifsupp \else \input{appendix} \fi

\label{lastpage}
\end{document}


\label{supp.firstpage}
\pagerange{\pageref{supp.firstpage}--\pageref{supp.lastpage}}
\maketitle

\section{Description of the Procedure}
\label{sec:gothic.description}

\begin{table*}
\centering
\caption{List of representative \texttt{objID}s}
\label{table:reps}
\begin{tabular}{|c|c|c|p{0.5\linewidth}|c|}
\hline
\textbf{No.} & \texttt{objID}                                                                                                         & \textbf{Common Name}    & \multicolumn{1}{c|}{\textbf{Description}} & \textbf{Figure}          \\ \hline
1 & \href{http://skyserver.sdss.org/dr12/en/tools/explore/summary.aspx?ra=174.1212\&dec=+21.5964}{1237667734504407133}   & MRK 739                 & Standard example of DNG  & Main Figure 1                           \\ \hline
2 & \href{http://skyserver.sdss.org/dr15/en/tools/explore/summary.aspx?ra=070.78608\&dec=+00.74408}{1237666301638279613} & UGC 3141                & DNG with nuclei far apart. This example has been chosen to demonstrate the limits of \textit{GOTHIC}. & \ref{fig:rep2}                         \\ \hline
3 & \href{http://skyserver.sdss.org/dr12/en/tools/explore/summary.aspx?id=1237651752385708153}{1237651752385708153}      & 2MASX J09520344+0052159 & Single nucleus galaxy with a star in the foreground. Such samples are potential false positives. & \ref{fig:rep3} \\ \hline
4 & \href{https://skyserver.sdss.org/dr16/en/tools/explore/summary.aspx?id=1237648720143647179}{1237648720143647179} & SDSS J095203.44+005216.3 & The frame does not contain sufficiently bright objects which can be detected by \textit{GOTHIC}. Many \texttt{objID}s in the blind search corresponding to such frames. & \ref{fig:rep4} \\ \hline
5 & \href{https://skyserver.sdss.org/dr16/en/tools/explore/summary.aspx?id=1237650794609246465}{1237650794609246465} & SDSS J094333.74-010551.3 & Single nucleus galaxy. These are the most commonly encountered objects in a blind search. Hence it's important to verify that they are reliably identified. & \ref{fig:rep5} \\ \hline
6 & \href{http://skyserver.sdss.org/dr16/en/tools/explore/Summary.aspx?id=1237648720688971816}{1237648720688971816} & SDSS J111508.35-003125.0 & Star mislabelled as a galaxy in SDSS which we frequently encountered in the blind search. Hence it is a clear representative. & \ref{fig:rep6} \\ \hline
7 & \href{http://skyserver.sdss.org/dr12/en/tools/explore/summary.aspx?id=1237650796756926591}{1237650796756926591} & 2MASS J09450600+0034516 & Two galaxies in the same line of sight. Similar to representative 3, this is a potential source of false-positives. & \ref{fig:rep7} \\ \hline
\end{tabular} 
\end{table*}

\begin{figure}
     \centering
     \begin{subfigure}[b]{0.15\textwidth}
         \centering
         \includegraphics[width=\textwidth]{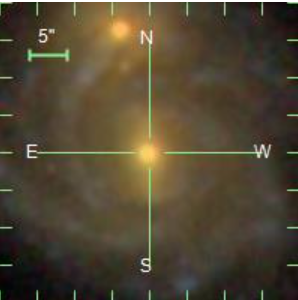}
         \caption{Representative 2}
         \label{fig:rep2}
     \end{subfigure}
     \hfill
     \begin{subfigure}[b]{0.15\textwidth}
         \centering
         \includegraphics[width=\textwidth]{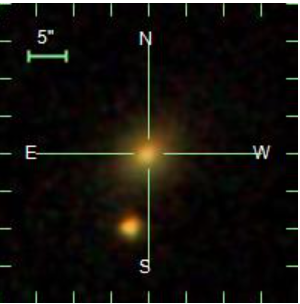}
         \caption{Representative 3}
         \label{fig:rep3}
     \end{subfigure}
     \hfill
     \begin{subfigure}[b]{0.15\textwidth}
         \centering
         \includegraphics[width=\textwidth]{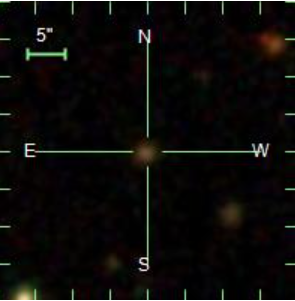}
         \caption{Representative 4}
         \label{fig:rep4}
     \end{subfigure}
     
     \begin{subfigure}[b]{0.15\textwidth}
         \centering
         \includegraphics[width=\textwidth]{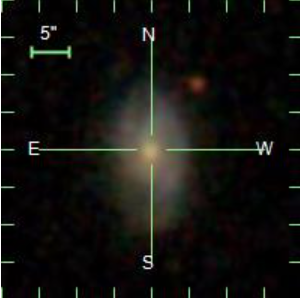}
         \caption{Representative 5}
         \label{fig:rep5}
     \end{subfigure}
     \hfill
     \begin{subfigure}[b]{0.15\textwidth}
         \centering
         \includegraphics[width=\textwidth]{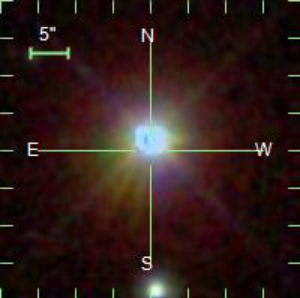}
         \caption{Representative 6}
         \label{fig:rep6}
     \end{subfigure}
     \hfill
     \begin{subfigure}[b]{0.15\textwidth}
         \centering
         \includegraphics[width=\textwidth]{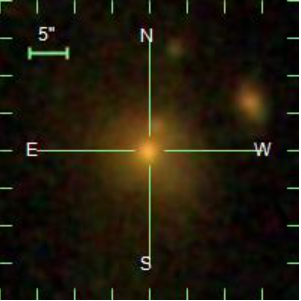}
         \caption{Representative 7}
         \label{fig:rep7}
     \end{subfigure}
    \caption{Representative sample of SDSS images seen in a blind search}
    \label{fig:reps}     
\end{figure}

We primarily describe the \gothic algorithm by means of example. In \tableref{reps}, we present a list of \objids from SDSS that are representative of the images of galaxies seen in practice. In the remainder of this document, we refer to the $n^{\text{th}}$ representative galaxy as "\mybf{Rep} $n$". The cutouts of each of these representatives are in \suppfigureref{reps} which displays \mybf{Rep} 2 through 7. \mybf{Rep} 1 is the archetypal double nuclei Galaxy MRK 739 which is shown in Main Figure 1. \footnote{The hyperlink to the image of Rep1 is unavailable in the table. Please refer to the main text}. 

We also provide algorithmic details of the sub-components of \gothic. For each sub-procedure below, we categorically state the representatives chosen for its demonstration along with grayscale images demonstrating the functionality of the sub-procedure. The coordinate axes of all grayscale images are scaled in pixels. Although we are interested in detecting galaxies only, we refer to the astronomical entity pointed to by an \objid as an ``object'' for ease of describing the \gothic algorithm.

\subsection{Image Normalization and Smoothing}
\label{sec:method.1}

\textbf{Representative Chosen} - 1 \\

\begin{figure}
     \centering
     \begin{subfigure}[b]{0.23\textwidth}
         \centering
         \includegraphics[width=\textwidth]{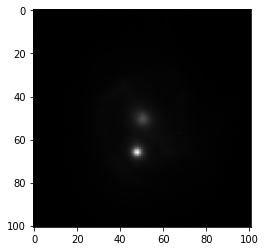}
         \caption{Raw cutout}
         \label{fig:rep1.cutout}
     \end{subfigure}
     \hfill
     \begin{subfigure}[b]{0.23\textwidth}
         \centering
         \includegraphics[width=\textwidth]{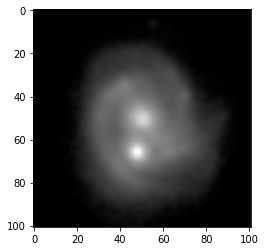}
         \caption{Normalization and Smoothening}
         \label{fig:rep1.logsmooth}
     \end{subfigure}
    \caption{The raw and normalized SDSS cutouts for Rep 1}
    \label{fig:raw.cutout}     
\end{figure}

Every object must be appropriately treated so that it is amenable for classification. Using \texttt{astropy} \citep{astropy}, a $40''$ cutout centred on an object's coordinates is performed. \mybf{Note that the angular size of \(40''\) typically corresponds to 100 pixels. Thus the cutout images are of size $100\times100$ in terms of pixels}. The raw cutout of \mybf{Rep} 1 is shown in \suppfigureref{rep1.cutout}. However, the light envelope of the galaxy and the dark background are not amenably contrasted for further analysis. The \mybf{nuclei at the centre of the galaxy are faint} and the body of the galaxy is not visible. 

To resolve this issue, we perform a log normalization of all pixel values in the cutout followed by appropriate scaling of the image within the range of $[0, 255]$. Subsequently, we perform \mybf{Gaussian} smoothing using a kernel with standard deviations of $(\sigma_x, \sigma_y)=(5, 5)$. The resulting image is shown in \suppfigureref{rep1.logsmooth}. 

\subsection{Bounding the Galaxy via Edge Detection and Convex Hull}
\label{sec:method.2}

\textbf{Representatives chosen} - 1, 5, 7 \\

Determining whether a galaxy is a DNG requires a reliable detection of two individual bright bulges within a galaxy envelope. Hence, it is important to determine the light envelope of the galaxy within which the search for dual peaks will take place. We use the Canny \citep{canny} edge detection technique, which utilizes the Sobel \citep{sobel} operator to create a discrete approximation of $\pdv{}{x}$ and $\pdv{}{y}$ at each pixel in the image, denoted by \(\Delta p_x, \Delta p_y\) respectively. Subsequently, the magnitude \(\sqrt{\Delta p_x^2 + \Delta p_y^2}\) is computed which represents how strongly a particular pixel is an edge. 

\begin{figure}
     \centering
     \begin{subfigure}[b]{0.23\textwidth}
         \centering
         \includegraphics[width=\textwidth]{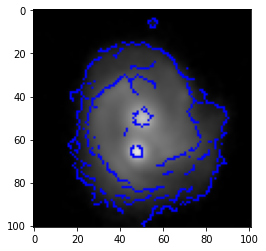}
          \caption{Rep 1}
         \label{fig:rep1.edges}
     \end{subfigure}
     \hfill
     \begin{subfigure}[b]{0.23\textwidth}
         \centering
         \includegraphics[width=\textwidth]{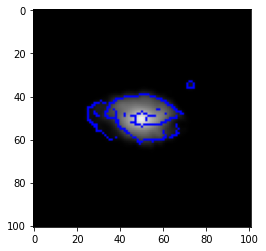}
         \caption{Rep 5}
         \label{fig:rep5.edges}
     \end{subfigure}
     \caption{Edges detected by Canny}
\end{figure}

For example, \suppfigureref{rep1.edges} and \suppfigureref{rep5.edges} show the edges detected by Canny (in blue) for Reps 1, 5 respectively. It is to be noted that Canny detects edges within the light envelope of the galaxy. This is expected since the Sobel operator measures the difference between neighboring pixel values at a given pixel coordinate. Generally, light profiles of galaxies in SDSS are not smooth and this causes sizeable differences between pixel values even within the light envelope which are mistakenly detected as edges. 

\begin{figure}
     \centering
     \begin{subfigure}[b]{0.23\textwidth}
         \centering
         \includegraphics[width=\textwidth]{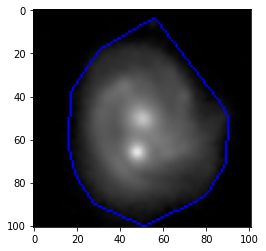}
         \caption{Rep 1}
         \label{fig:rep1.hull}
     \end{subfigure}
     \hfill
     \begin{subfigure}[b]{0.23\textwidth}
         \centering
         \includegraphics[width=\textwidth]{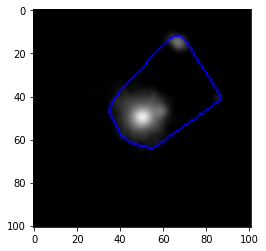}
         \caption{Rep 7}
         \label{fig:rep7.hull}
     \end{subfigure}
     \caption{Convex Hull of the Edges}
\end{figure}

We reiterate that the \mybf{Gaussian} Smoothing step in \sectionref{method.1} is necessary as it reduces the detection of false edges within the light envelope. If the smoothing step were skipped, the Canny algorithm would detect far more edges than necessary. On the other hand, excessive smoothing using higher values of $(\sigma_x, \sigma_y)$ would lead to no edges being detected. To resolve this issue, we find the convex hull of the pixel coordinates that have been detected as edges. The hulls for Reps 1 and 7 are shown in \suppfigureref{rep1.hull} and \suppfigureref{rep7.hull} respectively. Formally, the convex hull of a set of points $\mathcal{S}=\{(x,y)\}$ in 2-D is the smallest convex polygon that contains in its interior all points in $\mathcal{S}$. \citet{sklansky} and \citet{clrs} present an overview of the Convex Hull problem and efficient algorithms for computing it.

Note that the hull in \suppfigureref{rep7.hull} encloses a neighboring object. However, it is clear that this object is not of interest to us. In other words, we assume that the enclosed hull contains more objects that are of interest to us merely due to them being in the line of sight. Since our cutout size is $40^{\prime\prime}$, it is reasonable to expect the actual object referenced by the \objid would have many other objects in their line of sight and they need to be appropriately differentiated. Otherwise it can lead to unnecessarily high false-positive detections since two distinct objects that happen to be close together can be easily mistaken for a DNG. The sub-procedures in \sectionref{method.3}, \sectionref{method.4} and part of \sectionref{method.6} are dedicated to deal with such ``line of sight" scenarios.

\subsection{Numerical Fitting of the Galaxy Light Profile}
\label{sec:method.3}

\textbf{Representatives Chosen} - 1 \\

Since the convex hull might enclose a larger area than the object of interest, it needs to be ascertained what portion of the image within the hull resembles a galaxy envelope. In \suppfigureref{rep7.hull}, approximately one half of the hull (at the centre of the square image) contains the galaxy. This patch of light needs to be identified from the remainder of the hull. Next, we require a formal procedure for identifying the object of interest from the interior of the hull.

We make the assumption that the enclosed hull would always contain our galaxy of interest as well as other objects within the line of sight. The light profile of galaxies typically follow a distribution given by \citet{sersic}. This fact can be exploited to develop a procedure that separates a galaxy from the rest of the hull. Hence, we hypothesize that the pixel values in the hull will follow the \mybf{\sersic} distribution. The expression for the \mybf{\sersic} profile is as follows 
\begin{align}
    \log I(R) = \log I_0 - kR^{\frac{1}{n}} \label{eq:sersic.radius}
\end{align}
where $I(R)$ is the light intensity of a galaxy as a function of the radial distance $R$ from the galactic centre. The parameter $n$ is known as the \mybf{\sersic} index and it characterizes the overall shape of the light distribution. \citet{caon} states that $1 \leq n \leq 15$ fits the light profile of most galaxies. However, we have found that setting $0.25 \leq n \leq 15$ works better in practice.

\subsubsection{Radius to Frequency Relation}
\label{sec:rad.to.freq}

The \mybf{\sersic} profile is traditionally stated in terms of the radius from the galactic centre (\equationref{sersic.radius}). However, this information is not readily calculable and moreover, it is unsuitable for fitting to the data available at hand \ie the image histogram. As a simple heuristic, we use the taxicab norm of the pixel coordinates to represent the radial distance from the galactic centre. The taxicab norm \(R\) of the point \((x,y)\) is \(R=|x|+|y|\).

In order to translate the taxicab radius $R$ (which appears as the independent variable in the \mybf{\sersic} equation) into frequency $f$ (which is the data available to us). To find the relation from \(R\rightarrow f\), we need to find the number of distinct points \((x,y)\) that satisfy \(|x|+|y|=R\) for a given \(R\). It can be shown that the relation is \(f=4R\).

\begin{figure}
    \centering
    \includegraphics[width=\linewidth]{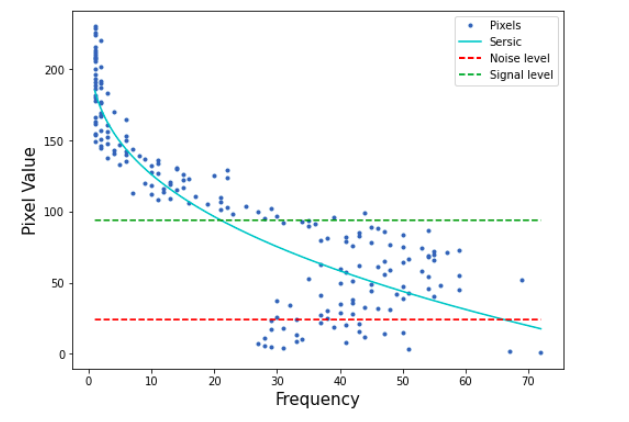}
    \caption{Rep 1 Sersic Fit}
    \label{fig:rep1.fit}
\end{figure}
\subsubsection{Fitting the Image Histogram}

We denote the histogram function of the smoothed image as $F\equiv F(p)$ which translates pixel values to their corresponding frequency values. Let the inverse histogram be $P\equiv P(f)$ that translates frequency numbers to pixel values of the smoothed image. The FITS files in SDSS report pixel values in a unit of flux called nanomaggies. It is these raw pixel values that are log-normalized and hence $P(f)$ would be in the form of (\equationref{sersic.radius}) apart from a constant additive factor. Hence, the $P(f)$ would have the form
\begin{align}
    P(f;k,n) = P_{max} - k\left(\frac{f}{4}\right)^{\frac{1}{n}} \label{eq:sersic.freq}
\end{align}
The term $P_{max}$ is the largest pixel value within the hull. The form of (\equationref{sersic.freq}) implicitly assumes that the pixel with the maximum value belongs to our object of interest and not a neighboring object. However, in practice if this assumption were not true, it would not affect the result significantly. This is because (\equationref{sersic.freq}) is numerically fit to the histogram data from the hull and the process of numerical fitting averages over the statistical noise in the data. Such errors would be present even if our assumption of the maximum pixel value belonging to our object of interest were not true. In other words, from the point of view of numerical fitting, the fact that the assumption is false can simply be interpreted as a form of statistical noise; hence it does not affect the final outcome.

$P(f)$ is numerically fitted to the observed histogram to obtain the best-fit values for its parameters $k, n$. We use the \texttt{optimize.curve\_fit} function from \texttt{scipy} \citep{scipy} to perform the numerical fitting. We show the fitted curve to the observed histogram of \mybf{Rep} 1 in \suppfigureref{rep1.fit}. The blue dots are the observed pixel values from the smoothed image and the teal line is the numerically fitted curve $P(f)$.

\begin{figure}
     \centering
     \begin{subfigure}[b]{0.23\textwidth}
         \centering
         \includegraphics[width=\textwidth]{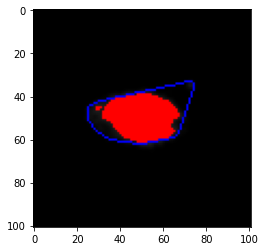}
         \caption{Rep 5}
         \label{fig:rep5.search}
     \end{subfigure}
     \hfill
     \begin{subfigure}[b]{0.23\textwidth}
         \centering
         \includegraphics[width=\textwidth]{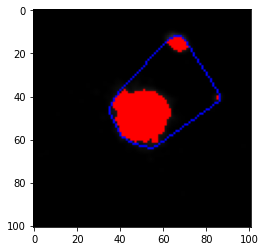}
         \caption{Rep 7}
         \label{fig:rep7.search}
     \end{subfigure}
     \caption{Search Region}
\end{figure}

\subsection{Determination of the Search Region}
\label{sec:method.4}

\textbf{Representatives Chosen} - 1, 5, 7 \\

After the curve $P(f)$ has been fitted to the histogram, it remains to be determined what pixel value separates the galaxy envelope from the background. This pixel value is known as the \emph{cutoff level}. It is also required in the ambient noise calculation (\sectionref{method.5}). With respect to the function $P(f)$, the pixel value whose corresponding frequency value captures $95\%$ of the total integrated light is chosen as the cutoff level. 

Mathematically, a typical $P(f)$ curve is a function of $f$ from $0$ to some maximum frequency value $f_{max}$. Thus, we define the following function
\begin{align}
    I_{integrated}(f) = \int_0^f P(f') \dd{f'}
\end{align}
The total integrated light would be
\begin{align}
    I_{total} = I_{integrated}\left(f_{max}\right)
\end{align}
As per our definition, The cutoff level would then be $P(f_{95})$ where $f_{95}$ satisfies
\begin{align}
    I_{integrated}\left(f_{95}\right) = 0.95\times I_{total}
\end{align}
We also define a \emph{signal} level as $P(f_{50})$. The definition of $f_{50}$ is analogous to that of $f_{95}$ above. The signal level is not used in determining the search region, but is crucial in the final classification phase (\sectionref{method.6})

With reference to \suppfigureref{rep1.fit}, the red line corresponds to the cutoff level and the green line is the signal level. All integral values mentioned are computed by numerically integrating the fit histogram $P(f)$ using the composite trapezoidal rule \citep{numint} whilst dividing the range of $[0, f_{max}]$ into $10,000$ steps.

\suppfigureref{rep5.search} and \suppfigureref{rep7.search} show in red the region above the cutoff level. This is the \emph{search region} where we look for dual peaks of light that characterizes DNGs. Note that the search region is included within the hull, and it perfectly masks the galaxy envelope (compare with \suppfigureref{rep5} and \suppfigureref{rep7} for reference). However, neighboring objects are mistakenly included at times within the envelope since their pixel values are above the cutoff (\suppfigureref{rep7.search}). The techniques in \sectionref{method.6} handle such scenarios.

\subsection{Iterated Hill Climbing}
\label{sec:method.5}

\textbf{Representatives chosen} - 1, 2, 3 \\

\begin{codebox}
\Procname{\proc{Hill-Climb}($\id{pt}$, $\id{reg}$, $\id{img}$)}
\label{algo:hill.climb}
\li \id{newpt} $\gets$ \texttt{NULL}
\li \While \id{pt} $\neq$ \id{newpt}
\li \Do \id{neighs} $\gets$ \proc{Get-Neighbors}($\id{pt}$, 3, $\id{reg}$)
\li \id{newpt} $\gets$ \id{pt}
\li \For \id{n} \textbf{in} \id{neighs}
\li \Do \If \id{img[n.x, n.y]} > \id{img[newpt.x, newpt.y]}
\End
\End
\li \id{pt} $\gets$ \id{newpt}
\End
\li \Return \id{newpt}
\End
\end{codebox}

\begin{codebox}
\Procname{\proc{Iterated-Hill-Climb}($\id{reg}, \id{img}$)}
\label{alg:iter.hill.climb}
\li \id{peak\_list} $\gets \Phi$ 
\li \id{i} $\gets$ 0
\li \Repeat \id{pt} $\gets$ \proc{Choose-Random}(\id{reg})
\li \id{peak} $\gets$ \proc{Hill-Climb}(\id{pt}, \id{reg}, \id{img})
\li \id{c_1} $\gets$ \id{peak} \textbf{not in} \id{peak\_list}
\li \id{c_2} $\gets$ \proc{snr}(\id{peak}) $\geq$ 3
\li \id{neighs} $\gets$ \proc{Get-Neighbors}($\id{pt}$, 3, $\id{reg}$)
\li \id{c_3} $\gets$ \proc{Intersection}(\id{neighs}, \id{peak\_list}) \textbf{is} $\Phi$
\li \If \id{c_1} \textbf{and} \id{c_2} \textbf{and} \id{c_3}
\li \Then \proc{Append}(\id{peak}, \id{peak\_list})
\End
\li \id{i} $\gets$ \id{i} + 1
\End
\li \Until \id{i} = $1,000$
\li \Return \id{peak\_list}
\end{codebox}

Having identified the galaxy envelope, we need to test for the presence of dual peaks of light which is characteristic of a DNG. However, a robust peak finding algorithm is first essential. We used a variant of the traditional hill-climbing algorithm to find the peaks. The idea is to start at a random pixel in the search region and then find the neighboring pixel which has the greatest pixel value. This process is repeated until no neighbor with a higher pixel value can be found. The final pixel at the end of this iterative process is reported to be the peak, provided the pixel value of the peak is greater than the \emph{signal} level (\sectionref{method.4}) and the SNR ratio at the  peak is $>3$. 

\subsubsection{Noise Level Inference for SNR Calculation}
\label{sec:noise.level.inference}

To determine the noise level in the SNR calculation, we averaged over a randomly chosen $3\times3$ grid of pixel values outside the convex hull. It needs to be ensured however that all constituent pixels of this chosen grid has pixel values below the cutoff level. This is done so that unrelated emissions that fall within the cutout do not erroneously influence the noise level inference. 

In Main Section 4.2, we have provided justifications for using a spectroscopic sample. Many objects in SDSS that are classified as a ``GALAXY'' are actually foreground stars. These objects contaminate our initial blind sample, and thus to avoid them, we have chosen a spectroscopic sample. It is worthwhile to note how the noise inference would take place for such \emph{contaminant} objects. They tend to have large diffraction spikes that permeates the entire region of the cutout. Hence, there may not be any $3\times3$ grid in the image that is below the noise level. Moreover, performing fitting a \mybf{\sersic} profile (\sectionref{method.4}) to a star is not scientifically sound (\mybf{\sersic} light profiles are meant to fit to spiral/elliptical galaxies). In other words, contaminant foreground stars typically have cutout images whose features are not amenable to numerical processing by \gothic. 

\subsubsection{Hill Climbing}
\label{sec:hill.climb}

Although we have performed \mybf{Gaussian} smoothing prior to the hill climbing, no image is ideal and there are residual non-uniformities and noise. As a result, the peak found by hill climbing is sensitive to the randomly initialized starting point \ie different starting points may lead to different peaks. To eliminate such uncertainties, we repeat $1,000$ instances of the hill-climbing subroutine initialized at random starting points and enlist all peaks thus found. 

\suppfigureref{rep1.hilllst} and \suppfigureref{rep2.hilllst} show how one random point ascends up to the peak via hill climbing. In both cases, the initialized point is near the edge of the galaxy envelope and takes small steps until it reaches the peak at one of the nuclei. There were 10 steps taken to reach a peak in the former, and 9 steps in the latter; each step has been marked by a red pixel. \suppfigureref{rep2.hillpeaks} and \suppfigureref{rep3.hillpeaks} show the peaks (marked in red) obtained by repeating hill climbing process for Reps 2, 3 respectively. \mybf{Rep} 2 (\suppfigureref{rep2.hillpeaks}) has peaks scattered across the envelope due to residual non-uniformities in the smoothed image. Although we perform $1,000$ hill climbs, not all peaks are stored (only the unique peaks are stored). Next we specify these conditions and provide the pseudocode for the hill climbing sub-routine. 

The formal hill climbing algorithm is given in the algorithm $\proc{Hill-Climb}$. It takes as its argument the randomly initialized starting point $\id{pt}$. $\id{reg}$ is list $\{(x,y)\}$ of pixel coordinates that represents the search region for peak finding (for example, the red region in \suppfigureref{rep5.search}). Finally, the variable $\id{img}$ is a dictionary that maps coordinates $(x, y)$ of the smoothed image to its corresponding pixel value (between 1 and 255 for grayscale images). $\texttt{NULL}$ represents a dummy value used for initialization. 

The procedure $\proc{Get-Neighbors}$ returns a list of pixels upto the nearest $3^{\text{rd}}$ neighbors (in terms of the supremum norm) of the input point $\id{pt}$. These neighbors form a $7\times7$ box at the centre of which lies the point $\id{pt}$. Variable $\id{reg}$ is also required by the procedure so that it can exclude any point in the $7\times7$ box that may lie outside the search (red) region. 

Semantically, the variable $\id{newpt}$ represents the coordinate that has the highest pixel value in comparison to the current $\id{pt}$. Its value is determined by lines 4 to 7 by iterative comparison of pixel values with each of the neighbors of $\id{pt}$ using the image dictionary $\id{img}$. If no higher neighbor is found, we obtain the condition $\id{pt}=\id{newpt}$. This is to be interpreted as --- \emph{no higher neighbor of $\id{pt}$ was found, so it must be a peak}. Finally, the while loop exits due to the guard condition in line 2 and it is returned in line 9.

It is possible to consider only the $1^{\text{st}}$ or $2^{\text{nd}}$ nearest neighbors when performing the neighbor search, however, it leads to a longer trail between the initial point and the final peak due to smaller discrete steps. This results in a slowdown of the $\proc{Hill-Climb}$ procedure. For optimal performance, we consider $3^{\text{rd}}$ neighbors only. Using higher-order neighbors such as $4, 5$ may lead to false negatives in the scenarios where the DNG peaks are in close proximity \ie the highest peak would always be chosen if it happens to be a close neighbor of the second highest peak. Hence, it could be mistakenly reported as a single nucleus galaxy.

\subsubsection{Iterated Hill Climbing}
\label{sec:iter.hill.climb}

Algorithm \ref{alg:iter.hill.climb} is the iterated hill climbing algorithm that is used to find the list of candidate peaks. It takes as input the search region $\id{reg}$ and the image dictionary $\id{img}$. Line 1 initializes the variable $\id{peak\_list}$ to the empty list $\Phi$ and the counter $\id{i}$ is set to $0$ in line 2.

It first chooses a random point in the search region via procedure $\proc{Choose-Random}$ at line 3, and it is subsequently passed to the hill-climbing subroutine in line 4. After a peak has been found, it is subjected to the following elimination conditions to decide if it should be included in $\id{peak\_list}$
\begin{enumerate}
    \item The SNR at the peak must be greater than 3 standard deviations
    \item The peak should not already exist in the list of peaks
    \item No neighboring pixel (upto the nearest third) must already exist in the list of peaks --- This condition has been included because $\proc{Hill-Climb}$ uses neighbors upto the nearest third in finding the highest neighboring pixel. Hence, it does not make sense to have two peaks in the peak list that are atmost third neighbors apart. This scenario is only relevant when any two peaks are equal in pixel value and happen to lie in each others' neighborhoods.
\end{enumerate}

\begin{figure}
     \centering
     \begin{subfigure}[b]{0.23\textwidth}
         \centering
         \includegraphics[width=\textwidth]{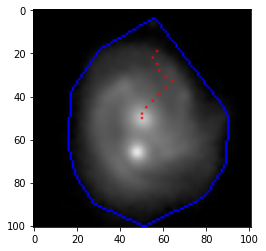}
         \caption{Rep 1}
         \label{fig:rep1.hilllst}
     \end{subfigure}
     \hfill
     \begin{subfigure}[b]{0.23\textwidth}
         \centering
         \includegraphics[width=\textwidth]{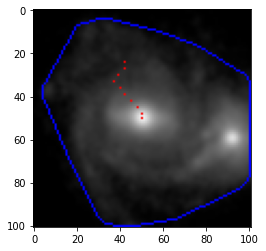}
         \caption{Rep 2}
         \label{fig:rep2.hilllst}
     \end{subfigure}
     \caption{Trace of Hill Climb}
\end{figure}

Only if the above three conditions are satisfied is the new peak appended to the list in line 10. Finally, the peak list is returned after $1,000$ iterations are concluded. It is precisely this peak list that has been marked in red in \suppfigureref{rep1.hilllst} and \suppfigureref{rep2.hilllst} respectively. As mentioned previously, \suppfigureref{rep2.hillpeaks} has multiple reported peaks within the galaxy envelope due to non-uniformity. Moreover, \suppfigureref{rep3.hillpeaks} has two peaks which clearly belong to two distinct objects and such cases are a potential source of false positives. Next, we present the techniques to handle such cases.

\subsubsection{Connected Components}
\label{sec:connected}

Simply stated, the number of connected components is the number disjoint portions of the search region of a galaxy. For example, there is 1 connected component in \mybf{Rep} 5 (\suppfigureref{rep5.search}) and 2 connected components in \mybf{Rep} 7 (\suppfigureref{rep7.search}). The algorithm for computing disjoint regions is a standard one and can be easily found in \citet{clrs}.

The astute viewer may point out that \mybf{Rep} 5 (\suppfigureref{rep5.search}) actually contains 2 connected components and that \mybf{Rep} 7 (\suppfigureref{rep7.search}) contains 3 connected components. Although it is true, these components are too small in size to represent any astronomical object (star or galaxy) in SDSS. It is an artefact of the noise in the image. Hence, all components that are \emph{smaller} than a certain threshold size are ignored.

\subsubsection{Most Probable Component}
\label{sec:probable}

\begin{figure}
    \centering
     \begin{subfigure}[b]{0.23\textwidth}
         \centering
         \includegraphics[width=\textwidth]{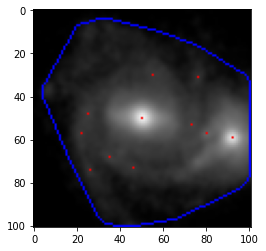}
         \caption{Rep 2}
         \label{fig:rep2.hillpeaks}
     \end{subfigure}%
     \begin{subfigure}[b]{0.23\textwidth}
         \centering
         \includegraphics[width=\textwidth]{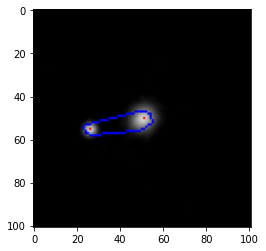}
         \caption{Rep 3}
         \label{fig:rep3.hillpeaks}
     \end{subfigure}
     \caption{Peaks for $1000$ instances of Hill Climb}
\end{figure}

When there are multiple peaks dispersed across multiple components, it is crucial to identify which component contains our peak(s) of interest. In a nutshell, a component $\mathcal{C}$ is a set of pixel coordinates $\mathcal{C}=\{(x,y)\}$. The number of pixels in a component would be the cardinality, $|\mathcal{C}|$. The search region for an image can be typically broken down into $n$ components as $\{\mathcal{C}_1, \mathcal{C}_2, \ldots, \mathcal{C}_n\}$. One of these components contains our peak of interest, denote it by $\mathcal{C}_p$. We describe intuitively the expected characteristics of such a $\mathcal{C}_p$ with the following heuristics. \\

\noindent\textbf{Heuristic (i) :} The component must be close to the centre of the image. This is intuitive as we take a cutout centred at the coordinates of an object. \\
\noindent\textbf{Heuristic (ii) :} The component must have a higher cardinality than other components. This ought to be true in most cases as other stray objects on the line of sight of the cutout are expected to be smaller in size. \\

It would be incorrect to choose a component that has higher cardinality but is distant from the centre of the image as it is more likely to represent a stray object on the line of sight. Hence heuristic (i) above must be given higher priority than heuristic (ii). 

To capture these notions into a mathematical formulation, we define the \emph{midpoint} of a component as the average of its constituent pixel coordinates \ie $m_{\mathcal{C}} = \left (\frac{\sum x}{|\mathcal{C}|}, \frac{\sum y}{|\mathcal{C}|}\right)$ for all $(x, y) \in \mathcal{C}$. We also define the \emph{distance} of component is its 2-D pixel distance from the centre of the image, $p$\footnote{
The $40''\times40''$ cutouts almost always correspond to $100$ pixel $\times100$ pixel images. Thus, the centre coordinate $p=(50,50)$
}. Formally stated, it is $d_{\mathcal{C}} = \sqrt{\lVert m_{\mathcal{C}} - p  \rVert^2}$. Next, we informally describe the steps to identify the most probable component $\mathcal{C}_p$. \\

\noindent\textbf{Step 1} - We aggregate the components into bins of 10 units of pixel distance. For example, all components with their distance $0\leq d_{\mathcal{C}} < 10$ would be put in the first bin, $10\leq d_{\mathcal{C}} < 20$ would be put in the second bin and so on. After the binning process, we consider only the first bin, $\mathcal{B}_1$, as all its constituent components are closest to the centre of the image. This captures heuristic (i). 

\noindent\textbf{Step 2} - We sort the constituents of $\mathcal{B}_1$ in descending order of their cardinality. Finally, we set the most probable component $\mathcal{C}_p$ as the first element of $\mathcal{B}_1$, which would be the largest component. This captures heuristic (ii). \\

This two step procedure gives more priority to heuristic (i) than (ii) because the binning procedure eliminates components that are far from the centre. The large component from only the \emph{first} bin $\mathcal{B}_1$ is chosen. 

\subsection{Final Classification}
\label{sec:method.6}

\begin{figure*}
    \centering
    \begin{subfigure}[b]{0.15\textwidth}
         \centering
         \includegraphics[width=\textwidth]{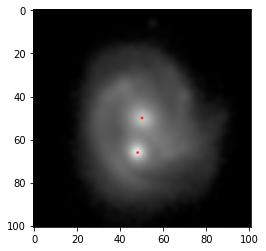}
         \caption{Rep 1}
         \label{fig:rep1.finpeaks}
     \end{subfigure}%
     \begin{subfigure}[b]{0.15\textwidth}
         \centering
         \includegraphics[width=\textwidth]{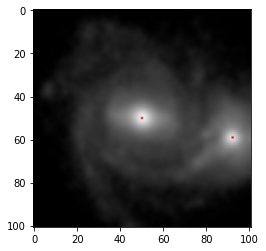}
         \caption{Rep 2}
         \label{fig:rep2.finpeaks}
     \end{subfigure}%
     \begin{subfigure}[b]{0.15\textwidth}
         \centering
         \includegraphics[width=\textwidth]{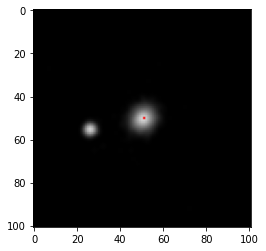}
         \caption{Rep 3}
         \label{fig:rep3.finpeaks}
     \end{subfigure}%
     \begin{subfigure}[b]{0.15\textwidth}
         \centering
         \includegraphics[width=\textwidth]{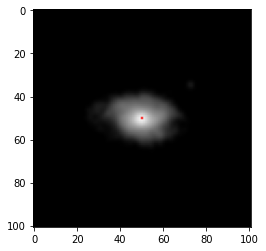}
         \caption{Rep 5}
         \label{fig:rep5.finpeaks}
     \end{subfigure}%
     \begin{subfigure}[b]{0.15\textwidth}
         \centering
         \includegraphics[width=\textwidth]{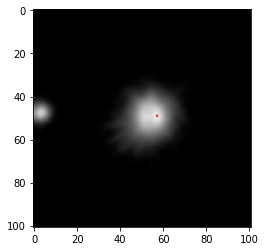}
         \caption{Rep 6}
         \label{fig:rep6.finpeaks}
     \end{subfigure}%
     \begin{subfigure}[b]{0.15\textwidth}
         \centering
         \includegraphics[width=\textwidth]{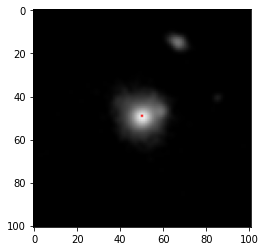}
         \caption{Rep 7}
         \label{fig:rep7.finpeaks}
     \end{subfigure}
     \caption{Final peaks after classification (marked as red pixel)}
     \label{fig:finpeaks}
\end{figure*}

\textbf{Representatives Chosen} - 2, 3 \\

To provide the final verdict for whether a galaxy is single or double nuclei, we use a simple criteria as follows

\noindent\textbf{Case 1} - If no peak is found by iterated hill climbing, the source is discarded. \\
\noindent\textbf{Case 2} - If one peak is found, then the galaxy is a single nucleus galaxy. \\
\noindent\textbf{Case 3} - If the number of peaks found is $\geq 2$, then we have two sub-cases based on how the peaks are distributed in the various \emph{connected-components} of the search region
\begin{enumerate}
    \item \textit{Subcase 3.1} - If the two brightest peaks are in the same component, then the galaxy is double nuclei. Otherwise it is judged as a single nucleus.
    \item \textit{Subcase 3.2} - The connected component with the highest \emph{chance} of representing a bonafide galaxy is selected. If it has one peak only, it is reported as a single nucleus. Otherwise, it is reported as double nuclei. 
\end{enumerate}

We justify each case of our classification criteria -

\noindent\textbf{Case 1} - It is possible that the Iterated Hill Climbing procedure returns no peak. This scenario can occur due to the first elimination criterion mentioned at the end of \sectionref{hill.climb}. \\
\noindent\textbf{Case 2} - This represents a non ambiguous detection of a single nucleus galaxy. \\
\noindent\textbf{Case 3} - The detection of more than one peak requires careful analysis. If two or more peaks are detected, naively adjudging them as double nuclei would lead to excess false positive detections. \suppfigureref{rep3.hillpeaks} is one such case where two peaks are detected, but they clearly belong to two distinct objects in the same field of view, and hence must be adjudged as a single nucleus galaxy. This \emph{intuitive} notion of distinct objects is captured by the \emph{mathematical} notion of connected components (\sectionref{connected}). Moreover, it is not enough to simply report that a galaxy is single/double nuclei. We wish to know the exact coordinates of the peak(s) detected. Thus, in the case where $\ge 2$ peaks are detected, it must be figured which two of the multiple peaks faithfully represent a double nuclei galaxy. \suppfigureref{rep2.hillpeaks} represents a scenario of multiple peaks in one connected component. In the general scenario, we would have multiple peaks detected in multiple connected components. To faithfully determine whether the galaxy is single/double nuclei, we require a notion of the \emph{most probable} connected component that is likely to contain the peak(s) of interest (\sectionref{probable}).

After the peaks have been located, we transform their pixel coordinates to astronomical coordinates (ra, dec) with the help of \texttt{astropy}. 

\subsubsection{Edge Cases in the Classification}
\label{sec:method.7}

We mention the edge cases that \gothic cannot handle reliably.
\begin{enumerate}
    \item If the Canny procedure cannot detect edges, the galaxy is not amenable to further processing and hence rejected.
    \item We expect that the centre of the cutout would be inside the convex hull formed from the edges reported by Canny. If this is not the case, the object (if any) at the centre of the image is not bright enough for further processing, and subsequently rejected.
    \item On rare occasions, \texttt{scipy} throws errors while fitting the \mybf{\sersic} profile to the pixel histogram. In other cases, it can time out as well. Such galaxies are rejected.
\end{enumerate}

\subsubsection{Performance of \gothic on the representatives}

We show the peaks detected by \gothic on the representative examples in \suppfigureref{finpeaks}. The final classification reported and peaks reported by \gothic agree well with the representative images in \suppfigureref{reps} and their corresponding descriptions in \tableref{reps}.

The two nuclei in the bonafide DNG examples (\mybf{Rep} 1 and \mybf{Rep} 2) are clearly identified. \mybf{Rep} 3 and \mybf{Rep} 7 showcase a star/galaxy in the line of sight of the central object respectively, and \gothic classifies both of them as a single galaxy. \mybf{Rep} 5 is a single nucleus galaxy with no object in its line of sight and is correctly classified as a single nucleus galaxy. \mybf{Rep} 6 is a star (mislabelled as galaxy in SDSS) and although it is reported to have one peak, it does not make a difference to our final aim of searching for DNGs from a blind sample (Main Section 4). \mybf{Rep} 4 is a faint galaxy and \gothic does not detect any peaks in it, and hence has been rightly excluded from \suppfigureref{finpeaks}.

\section{Performance of \gothic on the Gimeno Catalog}
\label{sec:verification.gimeno}

The pixel coordinates of the detected peak(s) when \gothic is run on the Gimeno Catalog and the control sample is shown in \tableref{gimeno.verification} and \tableref{control.verification} respectively. Each row in the table contains the \objid of the galaxy and the pixel coordinates of the detected peaks within the cutout.

\begin{table}
\caption{Results of \textit{GOTHIC} on the Gimeno Catalog - For each objID, the pixel coordinates of the detected peaks is given}
\label{table:gimeno.verification}
\centering
\begin{tabular}{|c|c|c|}
\hline
\texttt{objID} & \textbf{Peak 1} & \textbf{Peak 2} \\ \hline
1237678661428314227 & (50, 50) & (56, 41) \\ \hline
1237651737370296329 & (50, 51) & (36, 63) \\ \hline
1237661433243828392 & (61, 51) & (50, 50) \\ \hline
1237680265075884063 & (62, 32) & (50, 50) \\ \hline
1237678622230773769 & (50, 50) & (69, 56) \\ \hline
1237663530798219364 & (50, 49) & (38, 37) \\ \hline
1237680332175769723 & (49, 50) & (57, 40) \\ \hline
1237661950781685800 & (51, 32) & (50, 50) \\ \hline
1237668670253564040 & (50, 50) & (45, 39) \\ \hline
1237661850396327969 & (50, 50) & (21, 50) \\ \hline
1237680286536826907 & (49, 50) & (1, 33) \\ \hline
1237658801497178131 & (50, 50) & (43, 42) \\ \hline
1237679436129632275 & (50, 50) & (50, 58) \\ \hline
1237661816027938877 & (41, 44) & (50, 51) \\ \hline
1237655463774191765 & (93, 24) & (50, 49) \\ \hline
1237680529207066652 & (50, 50) & (50, 62) \\ \hline
1237665329853562899 & (57, 24) & (50, 49) \\ \hline
1237661063879000073 & (50, 50) & (69, 69) \\ \hline
1237658204517171275 & (50, 50) & (44, 25) \\ \hline
1237679456529743995 & (56, 58) & (46, 42) \\ \hline
1237664289930346555 & (53, 62) & (50, 48) \\ \hline
1237661950783979531 & (50, 50) & (60, 60) \\ \hline
1237680271501426770 & (50, 50) & (48, 37) \\ \hline
1237661358620082187 & (50, 50) & (57, 57) \\ \hline
1237653612113166459 & (50, 50) & (61, 53) \\ \hline
1237651539239895057 & (51, 50) & (45, 59) \\ \hline
1237657589780512816 & (50, 50) & (43, 42) \\ \hline
1237654400761790563 & (50, 50) & (59, 55) \\ \hline
1237650761852846175 & (50, 50) & (68, 59) \\ \hline
1237654382516699209 & (49, 50) & (49, 36) \\ \hline
1237655472898637896 & (50, 50) & (52, 40) \\ \hline
1237670961092558855 & (50, 50) & (28, 16) \\ \hline
1237657856073269409 & (50, 50) & (59, 27) \\ \hline
1237670961628971177 & (50, 61) & (49, 50) \\ \hline
1237666300016591063 & (50, 50) & (66, 59) \\ \hline
1237655463772618940 & (50, 50) & (34, 55) \\ \hline
1237665569292943466 & (50, 50) & (41, 49) \\ \hline
1237661977085018198 & (50, 50) & (68, 40) \\ \hline
1237678598094454843 & (50, 50) & (49, 45) \\ \hline
1237657771786960904 & (45, 59) & (50, 50) \\ \hline
1237664871897038979 & (50, 50) & (52, 40) \\ \hline
1237679581622698224 & (38, 47) & (49, 49) \\ \hline
1237654640211656710 & (33, 92) & (50, 50) \\ \hline
1237680245191868529 & (47, 51) & (51, 51) \\ \hline
1237658491207680027 & (74, 68) & (48, 51) \\ \hline
1237666301638279613 & (50, 50) & (59, 92) \\ \hline
1237667734504407135 & (54, 56) & (39, 59) \\ \hline
\end{tabular}
\end{table}
\begin{table}
\caption{Results of \textit{GOTHIC} on the Control Sample - For each objID, the pixel coordinates of the detected peak(s) is given}
\label{table:control.verification}
\centering
\begin{tabular}{|c|c|c|}
\hline
\texttt{objID} & \textbf{Peak 1} & \textbf{Peak 2} \\ \hline
1237648720142401611 & (50, 50) & - \\ \hline
1237650795146510627 & (50, 50) & - \\ \hline
1237650795146445031 & (50, 50) & - \\ \hline
1237648720142401670 & (50, 50) & - \\ \hline
1237648720142532891 & (50, 50) & - \\ \hline
1237650795146510903 & (30, 37) & - \\ \hline
1237648720142401774 & (40, 10) & - \\ \hline
1237650795146576091 & (50, 50) & - \\ \hline
1237650795146641532 & (50, 50) & - \\ \hline
1237650795146576030 & (50, 50) & - \\ \hline
1237650795146444862 & (9, 37) & - \\ \hline
1237650795146575935 & (28, 68) & - \\ \hline
1237651800158896649 & (60, 50) & (45, 50) \\ \hline
1237650794609574093 & (50, 50) & - \\ \hline
1237648720679469274 & (50, 50) & - \\ \hline
1237648721216274697 & (50, 50) & - \\ \hline
1237650795683512507 & (50, 50) & (40, 61) \\ \hline
1237650795683578031 & (50, 48) & (52, 53) \\ \hline
1237648721216340148 & (50, 50) & - \\ \hline
1237650795683578073 & (4, 45) & - \\ \hline
1237648721216405657 & (50, 50) & (52, 36) \\ \hline
1237648721216274720 & (41, 28) & - \\ \hline
1237650795683446969 & (9, 1) & - \\ \hline
1237650796220514489 & (50, 50) & (42, 45) \\ \hline
1237650795683512955 & (50, 50) & - \\ \hline
1237650795683512690 & (50, 50) & - \\ \hline
1237648720142336237 & (50, 50) & - \\ \hline
1237651799621959987 & (50, 50) & - \\ \hline
1237651799621894422 & (54, 50) & (55, 55) \\ \hline
1237650795146379400 & (50, 50) & - \\ \hline
1237648720142270656 & (50, 50) & - \\ \hline
1237650794609442930 & (50, 50) & - \\ \hline
1237648720142336185 & (50, 50) & - \\ \hline
1237650795683512717 & (19, 17) & - \\ \hline
1237650795146379447 & (50, 50) & - \\ \hline
1237651799621959813 & (64, 50) & (50, 50) \\ \hline
1237650795146379491 & (50, 50) & - \\ \hline
1237650795683446896 & (50, 50) & - \\ \hline
1237650795683446947 & (50, 50) & - \\ \hline
1237650795683447026 & (6, 10) & - \\ \hline
1237648720679338392 & (51, 50) & - \\ \hline
1237648720679338241 & (50, 50) & - \\ \hline
1237648720679338261 & (15, 14) & - \\ \hline
1237648720679338252 & (50, 50) & - \\ \hline
1237648721216209095 & (50, 50) & - \\ \hline
1237650795683446889 & (24, 48) & - \\ \hline
1237650796220252232 & (50, 50) & - \\ \hline
\end{tabular}
\end{table}

\bibliographystyle{mnras}
\bibliography{bibliography}

\label{supp.lastpage}